\documentclass[a4paper,12pt]{article}

\usepackage{graphicx}
\usepackage{array}
\usepackage{amsmath}
\usepackage{amssymb}
\usepackage{latexsym}
\usepackage{subfigure}
\usepackage{ulem}
\usepackage{color}
\renewcommand\sout{\bgroup \color{red} \ULdepth=-.5ex \ULset}

\newcommand {\beq}{\begin{eqnarray}}
\newcommand {\eeq}{\end{eqnarray}}
\newcommand {\non}{\nonumber\\}

\newcommand {\gto}{\stackrel{g}{\to}}

\newcommand {\1}[1]{\frac{1}{#1}}

\newcommand {\thb}{\bar{\theta}}

\newcommand {\ph}{\varphi}

\newcommand {\sig}{\sigma}

\newcommand {\Ph}{\Phi}
\newcommand {\Phd}{\Phi^{\dagger}}

\newcommand {\del}{\partial}
\newcommand {\dagg}{^{\dagger}}
\newcommand {\pri}{^{\prime}}
\newcommand {\prip}{^{\prime\prime}}

\newcommand {\prid}{^{\prime \dagger}}

\newcommand {\tr}{{\rm tr}\,}
\newcommand {\GC}{G^{\mathbb C}}
\newcommand {\HC}{H^{\mathbb C}}

\newcommand{\hs}[1]{\hspace{#1 mm}}

\makeatletter
\@addtoreset{equation}{section}
\makeatother

\date{empty}
\begin{document}
\begin{titlepage}
\null
\begin{flushright}
%arXiv:YYMM.XXXX
August, 2014
\end{flushright}
\vskip 0.5cm
\begin{center}
  {\Large \bf Higher Derivative Corrections to 
 \\
\vskip 0.3cm
Manifestly Supersymmetric Nonlinear Realizations
}
\vskip 1.1cm
\normalsize
\renewcommand\thefootnote{\alph{footnote}}

{\large
Muneto Nitta$^{\dagger}$\footnote{nitta(at)phys-h.keio.ac.jp}
and Shin Sasaki$^\ddagger$\footnote{shin-s(at)kitasato-u.ac.jp}
}
\vskip 0.7cm
  {\it
  $^\dagger$ 
Department of Physics, and Research and Education Center for Natural Sciences, \\
\vskip -0.2cm
Keio University, Hiyoshi 4-1-1, Yokohama, Kanagawa 223-8521, Japan  
\vskip 0.1cm
$^\ddagger$
  Department of Physics,  Kitasato University \\
  \vskip -0.2cm
  Sagamihara 252-0373, Japan
}
\vskip 0.5cm
\begin{abstract}
When global symmetries are spontaneously 
broken in supersymmetric vacua, 
there appear quasi-Nambu-Goldstone (NG) fermions 
as superpartners of NG bosons. 
In addition to these, there can appear quasi-NG bosons in general.
The quasi-NG bosons and fermions together with the NG bosons are
 organized into chiral multiplets.
K\"ahler potentials of low-energy effective theories were constructed 
some years ago as supersymmetric 
nonlinear realizations.
It is known that higher derivative terms in 
the superfield formalism 
often encounter the auxiliary field problem;
% the auxiliary fields are acted by
% space-time derivatives 
%and cannot be eliminated.
the auxiliary fields that accompanied with space-time derivatives and
it cannot be eliminated.
In this paper, we construct 
higher derivative corrections 
to supersymmetric nonlinear realizations
in the off-shell superfield formalism 
free from the auxiliary field problem.
As an example, we present the 
manifestly supersymmetric chiral Lagrangian.

\end{abstract}
\end{center}

\end{titlepage}

\newpage
\setcounter{footnote}{0}
\renewcommand\thefootnote{\arabic{footnote}}
\pagenumbering{arabic}
%%%%%%%%%%%%%%%%%%%%%%%%%%%%%%%%%%%%%%%%%%%%%%%%%
\section{Introduction}

Low-energy field theories 
can be described by only light fields 
when one integrates out massive particles 
above the scale which one considers.
In particular, when a global symmetry of 
Lagrangian or Hamiltonian is spontaneously 
broken in the ground state or vacuum, 
there appear Nambu-Goldstone (NG) bosons 
as massless scalar fields.
The low-energy dynamics of these NG bosons 
is solely determined from the symmetry argument.
When a symmetry group $G$ is spontaneously broken down to its subgroup $H$, 
the low-energy dynamics is governed
by a nonlinear sigma model whose target space is the 
 coset space $G/H$ \cite{Coleman:1969sm}.
A prime example is the chiral Lagrangian 
of pions which appear as NG bosons  
when the chiral symmetry of  QCD is spontaneously broken.  
Low-energy effective theories are usually expanded by 
the number of space-time derivatives, 
thereby they inevitably contain higher derivative corrections. 
It is known that the chiral perturbation theory
includes derivative corrections to the chiral Lagrangian \cite{Leutwyler:1993iq}. 

% On the other hand, 
% supersymmetry plays important roles 
% to control quantum corrections 
% in supersymmetric field theories, determining the exact low-energy
% dynamics \cite{Seiberg:1994rs}. 
On the other hand, supersymmetry plays important roles to control
quantum corrections in field theories and determines the exact
low-energy dynamics \cite{Seiberg:1994rs}.
It is also a necessary ingredient to define 
consistent string theories. 
It was also proposed 
as the most promising candidate to solve the naturalness problem in the 
Standard Model.
Among other things, 
when a global symmetry is spontaneously broken 
in supersymmetric vacua, 
there appear 
quasi-NG fermions \cite{Buchmuller:1982xn} 
in addition to the NG bosons.  
They are required to form chiral supermultiplets 
as superpartners of NG bosons. 
In model building of particle physics, 
quasi-NG fermions were identified as 
quarks in supersymmetric preon models
\cite{Buchmuller:1982tf}.  
The target spaces of supersymmetric nonlinear sigma models 
must be K\"ahler \cite{Zumino:1979et} 
because the lowest components of 
chiral superfields are complex scalar fields.
When a coset space $G/H$ is eventually K\"ahler, 
there are no additional massless fields.
However, $G/H$ is not K\"ahler in general, 
and in that case, 
 there must appear quasi-NG bosons \cite{Kugo:1983ma} 
in addition to the NG bosons, 
to parameterize a K\"ahler manifold.
In this case, target spaces of low-energy effective theories are
enlarged from $G/H$. 
In general, the problem to construct 
low-energy effective theories of massless fields 
reduces to finding $G$-invariant K\"ahler potentials.
The most general framework 
to construct 
$G$-invariant K\"ahler potentials 
was provided as
supersymmetric nonlinear 
realizations \cite{Bando:1983ab}.  
The authors of \cite{Bando:1983ab}
classified NG supermultiplets into 
P-type, containing two NG boson, 
and M-type, containing one NG boson and one quasi-NG boson.
In one extreme class
called a pure realization, 
all supermultiplets are of P-type and 
there are no quasi-NG bosons, 
which is possible only when $G/H$ 
happens to be K\"ahler.
In this case, 
the most general $G$-invariant K\"ahler potential 
up to K\"ahler transformations was constructed in Refs.~\cite{Bando:1983ab,Itoh:1985ha} 
(see Ref.~\cite{Bando:1987br} as a review), 
which is unique up to finite number of decay constants (K\"ahler class). 
This class was studied extensively in the literature 
(see, e.g.,~Refs.~\cite{KahlerG/H} 
and references in Ref.~\cite{Bando:1987br} ).
In the other extreme class 
called a maximal realization, 
all supermultiplets 
are of M-type 
so that there are the same number of quasi-NG bosons
with NG bosons. 
The target manifold in this case is 
a cotangent bundle $T^*(G/H)$,
whose cotangent directions are parameterized 
by quasi-NG bosons. 
For instance, the chiral symmetry breaking belongs to this 
class \cite{Kotcheff:1988ji}.
If there is at least one quasi-NG boson, 
the effective K\"ahler potential is an  
arbitrary function of strict $G$-invariants 
\cite{Bando:1983ab}.
Geometrically this arbitrariness corresponds to 
a degree of freedom to deform non-compact 
directions of 
the target space, 
which cannot be controlled by the isometry $G$ 
\cite{Kotcheff:1988ji, Shore:1988mn, Higashijima:1997ph, Nitta:1998qp}.
These directions are associated with the quasi-NG bosons.
It was proved that there must 
appear at least one quasi-NG boson 
in the absence of gauge interactions 
\cite{Lerche:1983qa,Shore:1984bh,Buchmuller:1986zp}. 
When there is a gauge symmetry on the other hand, 
pure realizations without quasi-NG bosons are possible 
by absorbing M-type superfields by the supersymmetric 
Higgs mechanism
\cite{Higashijima:1999ki}.

While the superfield formalism is 
one of the most powerful off-shell formulations to 
construct manifestly supersymmetric Lagrangians,
it often encounters an auxiliary field problem 
when 
higher derivative terms
exist in the Lagrangians. 
For example, 
chiral superfields with space-time derivatives
(e.g. $\partial_m \Phi$) contain
derivatives on the auxiliary fields $F$ 
so that they cannot be 
eliminated by their equations of motion. 
This problem was recognized 
\cite{Gates:1995fx,Nitta:2001rh} 
for a 
supersymmetric extension of Wess-Zumino-Witten (WZW) term 
\cite{Nemeschansky:1984cd} 
in the chiral Lagrangian of supersymmetric QCD.
A supersymmetric WZW term 
proposed in Ref.~\cite{Gates:2000rp} 
does not have this problem. 
Supersymmetric Lagrangians free from 
the auxiliary field problem were also known before,  
such as  
supersymmetric Dirac-Born-Infeld action \cite{RoTs}, 
supersymmetric higher derivative ${\mathbb C}P^1$ models \cite{BeNeSc,Fr}, 
supersymmetric baby Skyrme models \cite{Adam:2011hj,AdQuSaGuWe} 
and supersymmetric 
k-field theories \cite{AdQuSaGuWe2,AdQuSaGuWe3}. 
The most general model of chiral superfields 
with higher derivative terms was recently 
presented in Ref.~\cite{KhLeOv}, 
where it was called
a supersymmetric $P(X,\varphi)$ model. 
The higher derivative interaction can be written by using  
a target space tensor with two holomorphic and 
two anti-holomorphic indices which are both symmetric.
This term was first found in Ref.~\cite{Buchbinder:1994iw}
as a quantum correction term in 
a chiral model, 
and  the supersymmetric WZW term 
in Ref.~\cite{Gates:2000rp} 
also contains it \cite{Banin:2006db}.
The model in Ref.~\cite{KhLeOv}
was extended by 
the introduction of a superpotential \cite{SaYaYo} 
and coupling to supergravity \cite{KoLeOv, FaKe},
and was applied to the supersymmetric Galileon inflation models \cite{KhLeOv2} 
and the ghost condensation \cite{KoLeOv2}.
In our previous paper \cite{Nitta:2014pwa}, we have classified 
1/2 and 1/4 Bogomol'nyi-Prasad-Sommerfield (BPS) 
equations for domain walls, lumps, baby Skyrmions 
and domain wall junctions.
See also Ref.~\cite{Bolognesi:2014ova} for further study on baby Skyrmions.

In this paper, we  
construct higher derivative corrections 
to supersymmetric nonlinear realizations 
for spontaneous broken global symmetries 
with keeping supersymmetry. 
As the leading two derivative terms 
for pure realizations without quasi-NG bosons, 
we find that the higher derivative terms are unique 
up to constants.
On the other hand, 
higher derivative terms contain 
arbitrary functions 
in the presence of quasi-NG bosons. 
As one of the most important examples, 
we discuss chiral symmetry breaking in detail.

This paper is organized as follows.
In Sec.~\ref{sec:susy-nlr}, we give a brief review 
on supersymmetric nonlinear realizations. 
In Sec.~\ref{sec:hd-corr} 
we discuss higher derivative corrections 
to nonlinear realizations. 
In Sec.~\ref{sec:hdc}, 
we introduce the supersymmetric higher derivative
chiral model with four supercharges. 
We 
write down the equation of motion for the auxiliary fields
and analyze the structure of the on-shell Lagrangians. 
In Sec.~\ref{sec:pure}, we discuss 
higher derivative corrections to pure realizations 
in the absence of quasi-NG bosons, 
for which each massless chiral superfield contains two 
NG bosons and there are no quasi-NG bosons.
In Secs.~\ref{sec:hd-A-type} and \ref{sec:hd-C-type},
we discuss higher derivative corrections 
in the presence of quasi-NG bosons. 
In Sec.~\ref{sec:chisb}, 
we discuss 
higher derivative corrections 
for supersymmetric chiral symmetry breaking, 
which is a maximal realization 
where each massless chiral superfield contains one 
 NG boson and one quasi-NG boson.   
Section \ref{sec:conc} is devoted to conclusion and
discussions. 
We use the notation of the textbook of 
Wess and Bagger \cite{Wess:1992cp}.

%%%%%%%%%%%%%%%%%%%%%%%%%%%%%%%%%
\section{Supersymmetric Nonlinear Realizations: A Review}
\label{sec:susy-nlr}

In this section, we review supersymmetric 
nonlinear realizations formulated in Ref.~\cite{Bando:1983ab}.

%%%%%%%%%%%%%%%%%%%%%%%%%%%%%%%%%%%%%%%%%
\subsection{Global Symmetry Breaking in Supersymmetric Theories}
When a global symmetry 
group
$G$ is spontaneously broken down to its 
subgroup $H$, there appear massless Nambu-Goldstone (NG) bosons 
associated with 
broken generators of the coset manifold $G/H$. 
At low energies, interactions among these massless particles are 
described by the so-called nonlinear sigma models, 
whose Lagrangians 
in the leading order of derivative expansions 
are completely determined by the geometry of the 
target manifold $G/H$ parameterized by NG bosons 
as was found by Callan, Coleman, Wess and Zumino~\cite{Coleman:1969sm}.

In four-dimensional $\mathcal{N} = 1$ 
supersymmetric theories, scalar fields
belong to chiral superfields $\Phi^i$ 
$(i=1, \cdots, N)$ whose component expansion in the chiral base $y^m =
x^m + i \theta \sigma^m \bar{\theta}$ is 
\begin{align}
\Phi^i (y,\theta) = \varphi^i (y) 
+ \theta \psi^i (y) + \theta^2 F^i(y),
\end{align}
where $\varphi^i$ is the complex scalar field, $\psi^i$ is the Weyl
fermion and $F^i$ is the complex auxiliary field.

When a global symmetry is spontaneously broken 
in supersymmetric vacua, there appear massless fermions 
$\psi^i$ 
as supersymmetric partners of NG bosons~\cite{Buchmuller:1982xn}. 
These massless fermions together with NG bosons are described by 
chiral superfields.
Since chiral superfields are complex, 
the supersymmetric nonlinear sigma models 
are closely related to the complex geometry; 
their target manifolds, where fields 
variables take their values, must be K\"{a}hler 
manifolds~\cite{Zumino:1979et}. 
If the coset manifold $G/H$ itself happens to be a K\"{a}hler 
manifold, both real and imaginary parts of the scalar 
components of chiral superfields are NG bosons. 
If $G/H$ is not a K\"{a}hler manifold, 
on the other hand, there is at least 
one chiral superfield whose real or imaginary part 
is not a NG boson. 
This additional massless boson is called 
the {\it quasi-NG boson}~\cite{Kugo:1983ma}. 

We explain how quasi-NG bosons appear. 
The spontaneous symmetry breaking of 
a global symmetry $G$ in supersymmetric theories 
is caused by the superpotential $W$: 
the chiral superfields acquire the vacuum expectation values 
$v = \left<\varphi\right>$ as a result of 
the F-term condition 
$ \frac{\partial W}{\partial \varphi} = 0$.
Since the superpotential $W$ is {\it holomorphic} 
namely, it contains only chiral superfields, 
this condition is {\it invariant under 
the complex extension of $G$, namely, $\GC$.} 
Hence, if we define the {\it complex isotropy group} 
$\hat H (\subset \GC)$ by\footnote{
We use the calligraphic font for a Lie algebra 
corresponding to a Lie group.} 
\beq
 \hat H v = v, \quad \hat {\cal H} v = 0, 
\eeq
the target space parameterized by NG and 
quasi-NG
bosons can be written as a complex coset space:  
\beq
 M \simeq \GC/\hat H.
\eeq
In general, $\hat H$ is larger than $\HC$,
and it is decomposed as 
\beq
 \hat {\cal H} = {\cal H}^{\mathbb C} \oplus {\cal B},
\eeq
where ${\cal B}$ consists of non-hermitian generators 
$E \in \hat{\cal H}$
and is called (the subalgebra of) 
the {\it Borel subalgebra} in $\hat{\cal H}$~\cite{Bando:1983ab}.\footnote{ 
In the group level, 
$\hat H$ can be written as a semi-direct product 
of $\HC$ and the Borel subgroup $B$: 
$\hat H = \HC \wedge B$.
Here the symbol $\wedge$ denotes a semi-direct 
product.
If there are two elements of $\hat H$, 
$hb$ and $h\pri b\pri$, 
where $h,h\pri \in \HC$ and $b,b\pri \in B$, 
their product is defined as 
$(hb)(h\pri b\pri)= h h\pri (h^{\prime -1} b h\pri) b\pri 
= (h h\pri) (b\prip b)$, 
where $b\prip = h^{\prime -1} b h\pri \in B$~\cite{Bando:1983ab}.
It is, however, sufficient to consider only
 the Lie algebra in this paper.
\label{semi-direct}
}
%%%
\begin{itemize}
\item
As an example, let us consider a doublet 
$\phi=( \phi_1 , \phi_2)^T$ of $G=SU(2)$ and suppose 
that they acquire the vacuum expectation values 
$v=(1 , 0)^T$.
Since the raising operator 
$\sig_+= \1{2} (\sig_1+i\sig_2) = 
\left(\begin{array}{cc}
0 & 1 \cr 
0 & 0
\end{array}
\right)$ 
satisfies $\sig_+v=0$, it is the complex unbroken 
generator in $\hat{\cal H}$. 
On the other hand, $\sig_3$ 
and the lowering operator $\sig_- (= {\sig_+}\dagg)$ 
are the elements of 
the broken generators in ${\cal \GC} -\hat{\cal H}$.
\end{itemize}
%%%

The coset representative can be written as
\beq
 \xi(\Phi) = \exp(i \Phi \cdot Z) \in \GC/\hat H, 
 \quad Z \in {\cal \GC} -\hat{\cal H},
\eeq
where $Z$ are complex broken generators and 
$\Phi$ are NG chiral superfields generated by them. 
There are two kinds of broken generators: 
the hermitian broken generators $X$ and 
the non-hermitian broken generators $\bar E$: 
\beq 
{\cal \GC} -\hat{\cal H} = \{Z\} = \{X,\bar E\}.
\eeq
The NG superfields $\Phi$ corresponding to non-hermitian 
and hermitian generators are called  
{\it P-type} (or {\it non-doubled-type}) and 
{\it M-type} (or {\it doubled-type}) superfields, 
respectively~\cite{Bando:1983ab,Lerche:1983qa}. 
Note that there are as many non-hermitian broken 
generators $\bar E$ as non-hermitian unbroken generators $E$, 
since they are hermitian conjugate to each other. 
On a suitable basis, 
$\bar E$ and $E$ can be written as 
off-diagonal lower and upper half matrices respectively.
%%%
\begin{itemize}
\item
In the previous example where the representative of 
$\GC/\hat H$ is given by 
$\phi =\exp{i(\varphi_3\sig_3+\varphi\sig_-)}\cdot v$, 
$\varphi_3$ is a M-type and $\varphi$ is 
a P-type superfield. 
The non-hermitian broken generator 
$\bar E = \sig_-$ written as a lower half matrix is 
hermitian conjugate to 
the non-hermitian unbroken generator $E = \sig_+$ 
written as a upper half matrix.
\end{itemize}

The directions parameterized by quasi-NG bosons are non-compact, 
whereas those of NG bosons are compact.\footnote{
We use the word ``compactness'' in the sense of topology.
The kinetic terms of quasi-NG bosons have the same sign as 
those of NG bosons.
}
The scalar components of the M-type 
superfields
consist of a quasi-NG boson in addition to a NG boson, 
whereas those of the P-type superfields
consist of two genuine NG bosons. 
This can be understood as follows: 
note that, for each non-hermitian broken generator $\bar E$, 
there is a non-hermitian unbroken generator $E$.
Since the vacuum is invariant under ${\hat H}$, 
we can multiply 
the representative of the coset manifold
by an arbitrary element of ${\hat H}$ from the right. 
Hence, for any P-type superfield
$\Phi$ generated by a non-hermitian generator $\bar E$, 
there exists an element $\exp (i\Phi\dagg E) \in {\hat H}$
such that 
\beq
 \xi v 
  &=& \exp i (\cdots + \Ph \bar E + \cdots ) v \non
  &=& \exp i (\cdots + \Ph \bar E + \cdots ) 
          \exp (i\Phi\dagg E) v \non
  &=& \exp i (\cdots + \Re\Phi X_1 
         + \Im\Phi X_2 + O(\Phi^2) + \cdots ) v,\label{pure-NG}
\eeq
where we have used the Baker-Campbell-Hausdorff formula 
and defined two hermitian broken generators 
$X_1 = \bar E + E,\;\; X_2 = i(\bar E - E)$. 
Here $\Re$ and $\Im$ denote real and imaginary parts, 
respectively.
Therefore two scalar components of the P-type 
superfield
parameterize compact directions, 
and hence are considered NG bosons.
On the other hand, since any M-type 
superfield
is generated by an hermitian generator, 
there is no partner in $\hat{\cal H}$. 
Therefore its imaginary part of scalar component 
parameterizes a non-compact direction, 
and hence is considered to be a quasi-NG boson.
\begin{itemize}
\item
In our previous example, we can rewrite it as 
$\exp{i(\varphi_3\sig_3+\Re\varphi\sig_1+\Im\varphi\sig_2)}
\cdot v$ by multiplying an appropriate factor generated by 
$\sig_+$ for sufficiently small $|\varphi_3|$ and 
$|\varphi|$. The NG bosons parameterizing $S^3\simeq SU(2)$ 
are $\Re\varphi_3, \ \Re\varphi, \ \Im\varphi\ $, 
whereas $\Im\varphi_3$ is the quasi-NG boson parameterizing the 
radius of $S^3$. 
\end{itemize}

As a notation, we write the number of chiral superfields  
$N_{\Phi}$
parameterizing the target manifold as
\beq
 N_{\Ph} = N_{\rm M} + N_{\rm P},
\eeq
where the numbers of the M-type and P-type 
superfields 
are denoted by $N_{\rm M}$ and $N_{\rm P}$, respectively. 
The number of quasi-NG bosons is~\footnote{
We use `$\dim_{\bf C}$' for complex dimensions and `dim' 
for real dimensions.}
\beq
 N_{\rm Q} = N_{\rm M} 
 = 2\dim_{\bf C} (\GC/\hat H) - \dim (G/H)
 = \dim (G/H) - \dim B .
\eeq
Hence if there is as large Borel subalgebra as 
the number of NG bosons, $\dim B = \dim (G/H)$, 
there is no quasi-NG boson. 
This case is called the {\it pure realization} 
({\it total pairing} or {\it non-doubling}). 
On the other hand, 
if there is no Borel subalgebra, $\dim B = 0$, 
there appear as many quasi-NG bosons as NG bosons. 
This case is called the {\it maximal realization} 
(or {\it full-doubling}).
It is known that 
a pure realization cannot occur 
in the model with a linear origin 
without gauge symmetry~\cite{Lerche:1983qa,Shore:1984bh,Buchmuller:1986zp}. 
It was shown in Ref.~\cite{Lerche:1983qa} that 
a maximal realization occurs when 
a field belonging to a real representation obtains 
a vacuum expectation value or 
when NG boson part $G/H$ brought by 
a vacuum expectation value 
is a symmetric space.
In the presence of a gauge symmetry, 
pure realizations without quasi-NG bosons are possible,
since gauge fields absorb 
M-type superfields 
as a consequence of the supersymmetric 
Higgs mechanism \cite{Higashijima:1999ki}.  

%%%%%%%%%%%%%%%%%%%%%%%%%%
\subsection{$G$-invariant K\"ahler potentials}

The kinetic term in the 
effective Lagrangian is described by 
the K\"{a}hler potential $K(\Phi,\Phi\dagg)$ 
of NG chiral superfields 
\beq    
{\cal L} 
  = \int d^2\theta d^2\thb\, K(\Ph,\Phd) 
  = - g_{i\bar{j}}(\ph,\bar{\ph}) \del_{\mu}\ph^i \del^{\mu}\bar{\ph}^{\bar{j}} 
     + \mbox{(fermion terms)},
\label{kinetic_term}
\eeq
where we have eliminated 
the auxiliary fields
$F^i$ by its equation of motion and 
$g_{i\bar{j}} \equiv \frac{\partial}{\partial \varphi^i}
\frac{\partial}{\partial \bar{\varphi}^{\bar{j}}} K(\varphi, \bar{\varphi})$
is 
the K\"{a}hler metric.
Since the K\"{a}hler potential includes 
both chiral and anti-chiral superfields, 
the symmetry group
of the effective theory is still $G$, 
but not its complexification. 
Hence our goal is to construct $G$-invariant 
K\"{a}hler potentials of 
complex coset spaces $\GC/\hat H$.
Here the $G$-invariance means
\beq
 K(\Phi,\Phi\dagg) \gto K(\Phi\pri,\Phi^{\prime\dagger}) =
  K(\Phi,\Phi\dagg) + F(\Phi,g) + F^*(\Phi\dagg,g), 
  \label{G-transf.}
\eeq
where 
$F$ ($F^*$) is a (anti-)holomorphic function of $\Phi$
($\Phi^{\dagger}$) which depends on $g \in G$.
The latter two terms in Eq.~\eqref{G-transf.} 
disappear 
in the superspace integral $\int d^4 \theta$.~\footnote{
Here $F (\Phi,g)$ is called 
the {\it cocycle function}, which satisfies 
the {\it cocycle condition}, 
$F (\Phi,g_2 g_1) = F(g_2 \Phi,g_1) + F(\Phi,g_2)$.
}
Since the redefinition of the K\"{a}hler potential by 
adding holomorphic and anti-holomorphic functions 
is called the {\it K\"{a}hler transformation}, 
we denote that it is $G$-invariant under 
a K\"{a}hler transformation or quasi $G$-invariant 
if $F(\Phi,g)$ exists in 
Eq.~(\ref{G-transf.}).

\begin{figure}[h]
\begin{center}
\leavevmode
  \includegraphics[scale=.37]{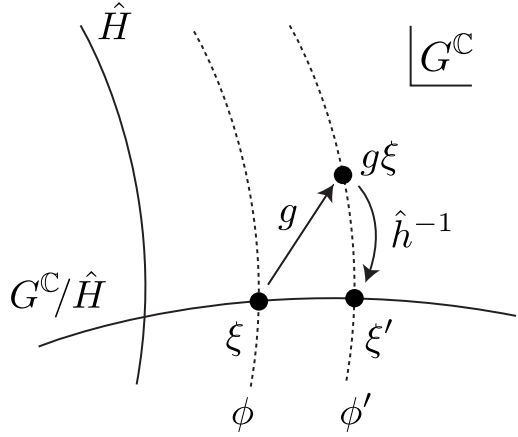} \\ 
\caption{
The $G$-transformation law for $\xi$.}
\label{fig:coset}
\end{center}
\end{figure}

First of all, we note that
the transformation law under $G$ of the representative $\xi$ 
of the complex coset $\GC/\hat H$ is
\beq
 \xi \gto \xi\pri = g \xi \hat h^{-1}(g,\xi),
\eeq
where $\hat h^{-1}(g,\xi)$ is a compensator 
to project $g \xi$ onto the coset representative, 
see Fig.~\ref{fig:coset}.
%%%%%%%%%%%%%%%% epsfig.sty %%%%%%%%%%%
%%%%%%%%%%%%%%%%%%%%%%%%%%%%%%%%%%%%%%%%

Bando {\it et}~al. constructed the following 
three types of $G$-invariant K\"{a}hler potentials called 
A-, B- and C-types \cite{Bando:1983ab}. 
%%%%%%%%%

{\bf A-type}. We prepare a representation $(\rho,V)$ of $G$ 
in the representation space $V$.
If there are $\hat H$ invariant vectors $v_a$, 
\beq
 \rho(\hat H) v_a = v_a,
\eeq
the transformation law of the quantity 
$\rho(\xi)v_a$ under $G$ is 
\beq
     \rho(\xi)v_a \stackrel{g}{\to} \rho(\xi ^{\prime})v_a 
 = \rho(g)\rho(\xi)\rho(\hat h^{-1})v_a
 = \rho(g)\rho(\xi)v_a.
\eeq
Then, by using strict $G$-invariants
\beq
  X_{ab} \equiv v^{\dagger}_a \rho(\xi^{\dagger}\xi)v_b,
 \label{eq:G-invariants}
\eeq
we can construct a $G$-invariant K\"{a}hler potential
\beq
&&  K_{\rm A} (\Phi,\Phi\dagg)
 = f( X_{ab} ),
\label{eq:A-type}
\eeq
where $f$ is an {\it arbitrary} real function 
of all possible $G$-invariants $X_{ab}$.

%\item
{\bf B-type}. It is sufficient to consider 
the fundamental representation~\cite{Bando:1983ab}, 
hence we do not write $\rho$ for simplicity.
We need the projection matrices, 
which project a fundamental representation space 
onto an $\hat H$ invariant subspace. 
They satisfy the projection conditions, 
\beq
 \eta_a \dagg = \eta_a, \hspace{1cm} 
 \eta_a \hat H \eta_a =\hat H \eta_a, \hspace{1cm} 
 \eta_a^2 = \eta_a.  \label{proj.}
\eeq
Define the projected determinant as
\beq 
 {\rm det}_{\eta} A \equiv \det (\eta A \eta + {\bf 1} -\eta), 
\eeq
where $\det_{\eta}$ 
stands for 
the determinant in the projected space. 
By using these, if we construct\footnote{
This can be rewritten as~\cite{BarMoshe:1994rx},
$K_{\rm B} 
 = \sum_a \log {\rm det}\pri (\xi \eta_a \xi\dagg)$,
where ${\rm det}\pri$ is a determinant 
except zero eigen values.
} 
\beq
 K_{\rm B} (\Phi,\Phi\dagg)
 = \sum_a c_a \log {\rm det}_{\eta_a} \xi\dagg \xi,
  \label{eq:kahler-coset}
\eeq
it is $G$-invariant up to a K\"{a}hler transformation:
\beq
\log {\rm det}_{\eta} \xi\dagg\xi 
 &\gto& \log {\rm det}_{\eta}
     \xi^{\prime \dagger}\xi\pri \non
&=& \log {\rm det}_{\eta}(\eta \xi\prid \xi\pri \eta)\non
&=& \log {\rm det}_{\eta}(\eta \hat h^{\dagger -1}
                   \xi\dagg \xi \hat h^{-1} \eta)\non
&=& \log {\rm det}_{\eta}(\eta \hat h^{\dagger -1} \eta
                 \xi\dagg \xi \eta \hat h^{-1}\eta )\non
&=& \log {\rm det}_{\eta}(\eta \hat h^{\dagger -1} \eta\eta
           \xi\dagg \xi \eta\eta \hat h^{-1} \eta)\non
&=& \log {\rm det}_{\eta} \xi\dagg \xi  
    + \log {\rm det}_{\eta} \hat h^{-1}
    + \log {\rm det}_{\eta} \hat h^{\dagger -1},
\eeq
where the last two terms include only chiral and anti-chiral 
superfields respectively, and disappear 
in the superspace integral $\int d^4 \theta$.~\footnote{
Here the cocycle function 
$F (\Phi,g) = \log {\rm det}_{\eta} \hat h^{-1}(g,\xi(\Phi))$ 
satisfies the cocycle condition.}
Here we have used Eq.~(\ref{proj.}). 

{\bf C-type}. Again, the fundamental representation is sufficient~\cite{Bando:1983ab}.
We define $[A]_{\eta}^{-1} \equiv
[\eta A \eta + {\bf 1} - \eta ]^{-1}$, 
where the inverse is calculated in the projected space. 
The quantities defined by\footnote{
The meaning of $P_a$ can be understood 
as follows~\cite{BarMoshe:1994rx}.
Since $P_a$ satisfies the properties
\beq
  P_a\dagg = P_a, \hs{10} 
  P_a^2 = P_a, \hs{10}
  \tr P_a = \tr \eta_a, \hs{10}
  P_a|_{\Phi = 0} = \eta_a, \nonumber
\eeq
$P_a$ can be considered to be the transformation of $\eta_a$ from 
the origin $\Phi = 0$ (or $\xi = 1$) to $\Phi \not=0$ in the manifold.
} 
\beq
  P_a = \xi \eta_a[ \xi\dagg \xi]_{\eta_a}^{-1}
        \eta_a \xi\dagg  \label{P_a}
\eeq
transform
 under $G$ as
\beq
  P \gto P\pri
&=&  \xi\pri\eta [ \xi\prid \xi\pri ]_{\eta}^{-1}
                 \eta \xi\prid \non
&=&  g \xi \hat h^{-1}\eta 
    [\eta \hat h^{-1 \dagger}\xi\dagg \xi  
        \hat h^{-1}\eta ]_{\eta}^{-1}
    \eta \hat h^{-1 \dagger} \xi\dagg g\dagg \non
&=&  g \xi \eta (\eta \hat h^{-1} \eta)
    [ \eta \hat h^{-1 \dagger} \eta \eta 
     \xi\dagg \xi \eta \eta \hat h^{-1} \eta]_{\eta}^{-1}
     (\eta \hat h^{-1\dagger} \eta)
      \eta \xi\dagg g\dagg \non
&=&  g \xi \eta [\hat h^{-1}]_{\eta}
    ( [\hat h^{-1\dagger}]_{\eta} [\xi\dagg \xi]_{\eta} 
      [\hat h^{-1}]_{\eta} )_{\eta}^{-1}
    [ \hat h^{-1\dagger}]_{\eta} 
     \eta \xi\dagg g\dagg \non
&=& g P g\dagg .
\eeq
By noting the relations 
\beq
 && P_a^2 
  = \xi \eta_a [ \xi\dagg \xi]_{\eta_a}^{-1}
    (\eta_a \xi\dagg \xi \eta_a)
    [ \xi\dagg \xi ]_{\eta_a}^{-1} \eta_a \xi\dagg  
 = P_a ,\\ 
 && \tr P_a 
  = \tr( [ \xi\dagg \xi ]_{\eta_a}^{-1}
       (\eta_a \xi\dagg \xi \eta_a) ) 
  = \tr \eta_a = {\rm const} ,
\eeq
a $G$ invariant K\"{a}hler potential can be constructed as 
\beq 
 K_{\rm C}(\Phi,\Phi\dagg) = f(\tr(P_a P_b),\; \tr(P_a P_b P_c),\cdots) ,
\eeq
where $f$ is again an {\it arbitrary} real function and 
all the indices $a, b, c, \cdots$ are different.

%%%%%%%%%%%%%%%%%%%%%%
\section{Higher Derivative Corrections} \label{sec:hd-corr}
In this section we study higher derivative corrections to supersymmetric
nonlinear realizations.
In the first subsection, we present 
general higher derivative chiral models 
with multiple chiral superfields. 
In the second subsection, we consider 
%higher derivative corrections to 
pure realizations described by 
B-type K\"ahler potentials,  
for which each massless chiral superfield contains two 
NG bosons.
In the third and fourth subsections, 
we consider A and C-type K\"ahler potentials, 
respectively, 
 for which some chiral superfields are M-type superfields, 
consisting of one quasi-NG boson and one genuine NG boson.

%%%%%%%%%%%%%%%%%%%%%%%%%%%%%
\subsection{Higher Derivative Chiral Models}\label{sec:hdc}

We consider higher derivative terms generated by multiple chiral
superfields $\Phi_i$ in which no dynamical (propagating) auxiliary fields exist.
The supersymmetric higher derivative term 
can be given by 
\cite{Adam:2011hj, AdQuSaGuWe, KhLeOv, Nitta:2014pwa}
\begin{align}
\mathcal{L}_{\rm H.D.}
=& 
 \frac{1}{16} \int \! d^4 \theta \ 
\Lambda_{ik\bar{j} \bar{l}}
(\Phi,
 \Phi^{\dagger}) 
D^{\alpha} \Phi^i
 D_{\alpha} \Phi^k \bar{D}_{\dot{\alpha}} \Phi^{\dagger \bar j}
 \bar{D}^{\dot{\alpha}} \Phi^{\dagger \bar{l}} .
\label{eq:Lagrangian}
\end{align}
Here the supercovariant derivatives are defined as
\begin{eqnarray}
D_{\alpha} = \frac{\partial}{\partial \theta^{\alpha}} + i
 (\sigma^m)_{\alpha \dot{\alpha}} \bar{\theta}^{\dot{\alpha}}
 \partial_m, \quad 
\bar{D}_{\dot{\alpha}} = - \frac{\partial}{\partial
\bar{\theta}^{\dot{\alpha}}} - i \theta^{\alpha} (\sigma^m)_{\alpha
\dot{\alpha}} \partial_m.
\end{eqnarray}
where the sigma matrices are $\sigma^m = (\mathbf{1}, \vec{\tau})$ 
with the Pauli matrices 
$\vec{\tau} = (\tau^1, \tau^2, \tau^3)$. 
Since the term $D_{\alpha} \Phi^i$ behaves as a vector
\beq 
  D_{\alpha} \Phi'^i = {\del \Phi'^i  \over \del \Phi^j }D_{\alpha} \Phi^j 
\eeq 
under field redefinition 
$\Phi^i \to \Phi^i{}'(\Phi)$, 
$\Lambda_{ik\bar{j} \bar{l}}$
can be regarded as  
a $(2,2)$ K\"ahler tensor 
symmetric in holomorphic and anti-holomorphic indices,   
whose components are 
functions of $\Phi^i$ and $\Phi^{\dagger \bar{i}}$
(admitting space-time derivatives acting on them).

We write down the bosonic components 
of the Lagrangian \eqref{eq:Lagrangian}.
The component expansion of the $\mathcal{N} = 1$ chiral superfield in
the $x$-basis is
\begin{equation}
\Phi^i (x, \theta, \bar{\theta}) = \varphi^i 
+ i \theta \sigma^m \bar{\theta} \partial_m \varphi^i + \frac{1}{4}
\theta^2 \bar{\theta}^2 \Box \varphi^i + \theta^2 F^i,
\end{equation}
where 
only the bosonic components are presented.
Then, 
the bosonic component of the supercovariant derivatives of $\Phi^i$ 
can be calculated as
\begin{align}
D^{\alpha} \Phi^i D_{\alpha} \Phi^k \bar{D}_{\dot{\alpha}}
 \Phi^{\dagger\bar{j}} \bar{D}^{\dot{\alpha}}
 \Phi^{\dagger\bar{l}} 
=& \ 16 \theta^2 \bar{\theta}^2 
\left[
\frac{}{}
(\partial_m \varphi^i \partial^m \varphi^k) (\partial_m
 \bar{\varphi}^{\bar{j}} \partial^m \bar{\varphi}^{\bar{l}})
\right. 
\notag \\
& 
\left.
- \frac{1}{2} 
\left(
\partial_m \varphi^i F^k + F^i \partial_m \varphi^k 
\right)
\left(
\partial^n \bar{\varphi}^{\bar{j}} \bar{F}^{\bar{l}} 
+ \bar{F}^{\bar{j}} \partial^n \bar{\varphi}^{\bar{l}}
\right)
+ F^i \bar{F}^{\bar{j}} F^k \bar{F}^{\bar{l}}
\right].  \label{eq:4th}
\end{align}
Since the bosonic part 
of the right hand side of \eqref{eq:4th} saturates
the Grassmann coordinate $\theta^2 \bar{\theta}^2$, 
only the lowest component of the tensor $\Lambda_{ik\bar{j} \bar{l}}$
contributes to the bosonic part of the Lagrangian.
Therefore the bosonic part of the Lagrangian 
\eqref{kinetic_term} 
with the higher derivative term \eqref{eq:Lagrangian} is 
\begin{align}
\mathcal{L}_b =& \ 
g_{i\bar j}
(- \partial_m \varphi^i \partial^m \bar{\varphi}^{\bar{j}} + F^i
 \bar{F}^{\bar{j}} )
+ \frac{\partial W}{\partial \varphi^i} F^i + \frac{\partial
 \bar{W}}{\partial \bar{\varphi}^{\bar{j}}} \bar{F}^{\bar{j}}
\notag \\
& + \Lambda_{ik\bar{j} \bar{l}} (\varphi, \bar{\varphi})
\left[
\frac{}{}
(\partial_m \varphi^i \partial^m \varphi^k) (\partial_n
 \bar{\varphi}^{\bar{j}} \partial^n \bar{\varphi}^{\bar{l}})
-
\partial_m \varphi^i F^k 
\partial^m \bar{\varphi}^{\bar{j}} \bar{F}^{\bar{l}} 
+ F^i \bar{F}^{\bar{j}} F^k \bar{F}^{\bar{l}}
\right],
\label{eq:comLagrangian}
\end{align}
where we have introduced the superpotential $W$ for generality.
The model is manifestly
(off-shell) supersymmetric and K\"ahler invariant provided that $K$ and $W$
are scalars and $\Lambda_{ik\bar{j} \bar{l}}$ is a tensor.
The auxiliary fields $F^i$ do not have space-time derivatives
and consequently can be eliminated by the following algebraic equation of motion, 
\begin{align}
g_{i\bar j}
F^i - 2 \partial_m \varphi^i F^k \Lambda_{ik\bar{j} \bar{l}} 
 \partial^m \bar{\varphi}^{\bar{l}} + 
2 \Lambda_{ik\bar{j} \bar{l}} F^i F^k \bar{F}^{\bar{l}} 
+ \frac{\partial \bar{W}}{\partial \bar{\varphi}^{\bar{j}}} 
= 0. \label{eq:af-eom}
\end{align}

Since NG fields are all massless, 
we consider the vanishing superpotential $W=0$\footnote{
If we consider the spontaneously breaking of approximate
symmetries, a non-zero superpotential $W$ that
provides small mass to the pseudo-NG modes is possible.
}.
In this case, $F^i = 0$ is a solution to this equation, 
and the on-shell Lagrangian becomes 
\begin{align}
\mathcal{L}_b = - 
g_{i\bar j}
\partial_m \varphi^i \partial^m
 \bar{\varphi}^{\bar{j}} 
+ \Lambda_{i k \bar{j} \bar{l}} 
(\partial_m \varphi^i \partial^m \varphi^k) (\partial_n \bar{\varphi}^{\bar{j}}
 \partial^n \bar{\varphi}^{\bar{l}}).
\label{on_shell_Lagrangian}
\end{align}
We call this canonical branch.
We note that the second term in \eqref{on_shell_Lagrangian} contains
more than the forth order of space-time derivatives for appropriate
functions $\Lambda_{ik\bar{j}\bar{l}}$. 
We will demonstrate an example of sixth-derivative terms in 
Sec.~\ref{sec:super-chiral-pert}.

In general, there are more solution other than $F^i = 0$, 
although 
an explicit solutions $F^i$ is not easy to find  
except for one component field. 
Indeed, for single superfield models, 
we have other on-shell branches associated with 
solutions $F^i \not= 0$ \cite{AdQuSaGuWe,Nitta:2014pwa}.
We call this non-canonical branch.
In the non-canonical branch, the ordinary kinetic term with two space-time
derivatives vanishes and the 
on-shell Lagrangian consists of only four-derivative 
terms.
Although it is interesting, we do not consider this branch 
because we are considering derivative expansions.

%%%%%%%%%%%%%%%%%%
\subsection{B-type (Pure Realizations)}\label{sec:pure}

When there are no quasi-NG modes, it is called a pure realization.
This is possible only when $G/H$ is eventually K\"ahler.
When there is a gauge symmetry, 
the pure realization without quasi-NG bosons is possible 
\cite{Higashijima:1999ki}.  
From the Borel's theorem, 
compact K\"ahler coset spaces $G/H$ can be written as
\beq 
G/H = G/[H_{\bf s.s.} \times U(1)^r]
\eeq  
with $H_{\rm s.s.}$ the semi-simple subgroup in $H$ and 
$r \equiv {\rm rank}\, G - {\rm rank }\, H_{\rm s.s.}$~\cite{Borel:1954}. 
In this case, there exists the isomorphism 
\beq 
  G/H \simeq \GC/\hat H.
\eeq 
The most general $G$-invariant K\"ahler potential 
(up to K\"ahler transformations) was 
shown to be written solely by B-type K\"ahler potentials 
and A and C-types were shown not to give independent 
K\"ahler potentials 
 \cite{Bando:1983ab,Itoh:1985ha,Bando:1987br}.

Now we consider higher derivative terms.
In this case, the problem is reduced to 
find $G$ invariant $(2,2)$ tensors $\Lambda_{ik\bar j \bar l}$ 
on the target manifold $G/H$.
The $G$-transformation on the fields are 
\beq
 \delta \Phi^i_A = k^{i}_A,
\eeq
where $k^{i}_A(\Phi)$ ($A = 1,2,\cdots,\dim G$) 
are holomorphic Killing vectors generated by the isometry $G$, preserving the metric 
${\cal L}_k g_{i\bar j} = 0$. 
The $(2,2)$ tensors $\Lambda_{ik\bar j \bar l}$ 
for higher derivative term 
must be preserved by the isometry $G$: 
${\cal L}_k \Lambda_{ik\bar j \bar l} = 0$. 
Then, $G$-invariant four derivative terms are given by
\beq
&& \mathcal{L}^{(4)} =  \frac{1}{16} \int \! d^4 \theta \ 
\Lambda_{ik\bar{j} \bar{l}}
(\Phi,
 \Phi^{\dagger}) 
D^{\alpha} \Phi^i
 D_{\alpha} \Phi^k \bar{D}_{\dot{\alpha}} \Phi^{\dagger \bar j}
 \bar{D}^{\dot{\alpha}} \Phi^{\dagger \bar{l}} 
, \non 
&&  \Lambda_{ik\bar j\bar l} 
= 
 w_1 g_{(i\bar j} g_{k\bar l)}
 + w_2 R_{i\bar jk\bar l} 
 + w_3 R_{(i\bar j} R_{k\bar l)} 
 + w_4  g_{(i\bar{j}} R_{k \bar{l})}
 \label{eq:chiral-4deriv-pure}
\eeq
where $R_{i\bar jk\bar l}$ and $R_{i\bar j}$ are 
the Riemann curvature and Ricci-form, respectively, 
brackets $(...)$ imply symmetrization 
over holomorphic and anti-holomorphic indices, 
and $w_{1,2,3}$ are real constants.
The scalar curvature $R$ is also 
invariant but it is just a constant for $G/H$.
The explicit form of the curvature tensor can be found 
in Ref.~\cite{Aoyama:2000vb}.
In some cases, 
the terms in Eq.~(\ref{eq:chiral-4deriv-pure}) 
are not independent. 
For Einstein manifolds, $R_{i\bar j} \sim g_{i \bar j}$ holds.
For instance, 
rank one cases $(r=1)$ belong to this class.

An important fact is that  
there are no strict $G$-invariant, 
unlike the case with quasi-NG bosons 
which we discuss in the next subsections. 
This is the reason why higher derivative 
terms are uniquely determined up to constants. 

As for derivative terms higher than four derivatives, 
one uses the covariant derivatives of tensors 
such as 
$D_{g} \bar D_{\bar h} R_{i\bar{j}k\bar{l}}$.
For instance, a six-derivative term can be constructed as
\beq
\mathcal{L}^{(6)} =  \frac{1}{16} \int \! d^4 \theta \ 
D_{g} \bar D_{\bar{h}}
 R_{i\bar{j}k\bar{l}} 
\del_m \Phi^g \del^m \Phi^{\dagger \bar h}
 D^{\alpha} \Phi^i
 D_{\alpha} \Phi^k 
 \bar{D}_{\dot{\alpha}} \Phi^{\dagger \bar j}
 \bar{D}^{\dot{\alpha}} \Phi^{\dagger \bar{l}} + \cdots.
\eeq

%%%%%%%%%%%%%%%%%%%%%%%%%%%%%
\subsection{A-type}\label{sec:hd-A-type}

The K\"ahler potential of A-type 
is given in Eq.~(\ref{eq:A-type}).
There are two ways to 
construct  $G$-invariant four-derivative terms 
using the A-type invariants. 
The first way is a geometrical method which is 
the same with pure realizations, 
and  the second way is a group theoretical method.

In the first method, 
$G$-invariant four-derivative terms are given by
\beq
&& \mathcal{L}^{(4)} 
=  \frac{1}{16} \int \! d^4 \theta \ 
\Lambda_{ik\bar{j} \bar{l}} (\Phi, \Phi^{\dagger}) 
 D^{\alpha} \Phi^i
 D_{\alpha} \Phi^k \bar{D}_{\dot{\alpha}} \Phi^{\dagger \bar j}
 \bar{D}^{\dot{\alpha}} \Phi^{\dagger \bar{l}} 
, \non 
&&  
\Lambda_{ik\bar j\bar l} 
=  
w_1( X_{ab} ) g_{(i\bar j} g_{k\bar l)} 
+ w_2( X_{ab}) R_{i\bar jk\bar l} 
+w_3( X_{ab} ) R_{(i\bar j} R_{k\bar l)} 
+ w_4 ( X_{ab}) g_{(i\bar{j}} R_{k \bar{l})} .
  \label{eq:chiral-4deriv-A}
\eeq
Unlike the B-type case, 
$w_{1,2,3,4}$ are {\it arbitrary functions}  
of the strict $G$-invariants $X_{ab}$. 
The scalar curvature $R$ is a function of 
$X_{ab}$ and is not included.

Now we introduce the second method 
 to construct $G$-invariant four-derivative terms. 
Here we do not write the representation $\rho$ for simplicity.
First, the Maurer-Cartan one form on 
$\GC / \hat H$ is given by 
\beq
 i \xi^{-1} d \xi =  (E^I_i (\Phi)X_I + \omega^a_i (\Phi)H_a) d\Phi^i
\eeq
with the holomorphic vielbein $E^I_i (\Phi)$ and 
the holomorphic connection $\omega^a_i (\Phi)$.
By using this expression, we calculate
\beq 
&& D_{\alpha} \xi = D_{\alpha} \Phi^i \del_i \xi
=   i \xi  (E^I_i (\Phi)X_I + \omega^a_i (\Phi)H_a)  D_{\alpha} \Phi^i ,\\
&& D_{\alpha} \xi v_a 
=  i  (\xi  X_I v_a) E^I_i (\Phi) D_{\alpha} \Phi^i.
\eeq
Then, the supercovariant 
derivatives of the $G$-invariants $X_{ab}$
given in Eq.~(\ref{eq:G-invariants}) can be calculated to be
\beq 
&& D_{\alpha} X_{ab}  
=     (v_a \xi^\dagger\xi  X_I v_b)  
 E^I_i (\Phi) D_{\alpha} \Phi^i \\ 
&& D^{\alpha} D_{\alpha} X_{ab}  
=     (v_a \xi^\dagger\xi X_J X_I v_b)  
 E^I_i (\Phi) E^J_j (\Phi) D^{\alpha} \Phi^i D_{\alpha} \Phi^j \\ 
&& \bar D^{\dot \alpha}  D_{\alpha} X_{ab}  
=     (v_a X_J^\dagger \xi^\dagger\xi  X_I v_b)  
 E^I_i (\Phi)  E^{*J}_j (\Phi^\dagger)
D_{\alpha} \Phi^i 
\bar D^{\dot \alpha} \Phi^{\dagger \bar j} ,\\ 
&& \bar D_{\dot \alpha}  D^{\alpha} X_{ab}  
\bar D^{\dot \alpha}  D_{\alpha} X_{cd}  
=    (v_a X_J^\dagger \xi^\dagger\xi  X_I v_b)  
      (v_c X_L^\dagger \xi^\dagger\xi  X_K v_d)  
 E^I_i (\Phi)  E^K_k (\Phi)  
  E^{*J}_j (\Phi^\dagger)
  E^{*L}_l (\Phi^\dagger)  \non
&& \hs{40} \times D^{\alpha} \Phi^i D_{\alpha} \Phi^k 
\bar D_{\dot \alpha} \Phi^{\dagger \bar j}
\bar D^{\dot \alpha} \Phi^{\dagger \bar l}, \\
&& 
\bar D_{\dot \alpha} 
\bar D^{\dot \alpha} 
D^{\alpha} D_{\alpha} X_{ab}  
=     (v_a  X_J^\dagger  X_L^\dagger \xi^\dagger\xi X_K X_I v_b)  
 E^I_i (\Phi) E^K_k (\Phi) E^{*J}_j (\Phi^\dagger)
  E^{*L}_l (\Phi^\dagger) \non
&& \hs{40} \times 
D^{\alpha} \Phi^i D_{\alpha} \Phi^k 
\bar D_{\dot \alpha} \Phi^{\dagger \bar j}
\bar D^{\dot \alpha} \Phi^{\dagger \bar l} , \\
&& 
D^{\alpha} D_{\alpha} X_{ab}  
\bar D_{\dot \alpha}  \bar D^{\dot \alpha} X_{cd} 
= (v_a \xi^\dagger\xi X_K X_I v_b)  
   (v_c  X_J^\dagger  X_L^\dagger \xi^\dagger\xi v_d)  
  E^I_i (\Phi) E^K_k (\Phi) E^{*J}_j (\Phi^\dagger)
  E^{*L}_l (\Phi^\dagger) \non
&& \hs{40} \times 
D^{\alpha} \Phi^i D_{\alpha} \Phi^k 
\bar D_{\dot \alpha} \Phi^{\dagger \bar j}
\bar D^{\dot \alpha} \Phi^{\dagger \bar l} .
\eeq 
By using these relations, 
four-derivative terms can be given by 
\beq 
&&\mathcal{L}^{(4)} 
=  \frac{1}{16} \int \! d^4 \theta \ 
\big[
 g_1^{ab} (X_{mn}) 
 \bar D_{\dot \alpha} \bar D^{\dot \alpha} 
 D^{\alpha} D_{\alpha} X_{ab} 
\non && \hs{28}
+ 
g_2^{abcd} (X_{mn})
 \bar D_{\dot \alpha}  D^{\alpha} X_{ab}  
 \bar D^{\dot \alpha}  D_{\alpha} X_{cd}  
+  
g_3^{abcd} (X_{mn})  
 D^{\alpha}   D_{\alpha} X_{ab}  
 \bar D_{\dot \alpha}  \bar D^{\dot \alpha} X_{cd}
\non &&   \hs{28}
+  
g_4^{abcdef} (X_{mn})  
 D^{\alpha} X_{ab}  D_{\alpha} X_{cd}  
 \bar D_{\dot \alpha}  \bar D^{\dot \alpha} X_{ef}
+  
g_5^{abcdef} (X_{mn})  
 D^{\alpha}  D_{\alpha}  X_{ab}    
 \bar D_{\dot \alpha}  X_{cd}   \bar D^{\dot \alpha} X_{ef}
\non &&  \hs{28}
 + 
g_6^{abcdef}  (X_{mn}) 
  D^{\alpha} X_{ab} D_{\alpha} \bar D_{\dot \alpha}  X_{cd} 
 \bar D^{\dot \alpha}  X_{ef} 
\non &&  \hs{28}
+ g_7^{abcdefgh}  (X_{mn}) 
  D^{\alpha} X_{ab} D_{\alpha} X_{cd} 
 \bar D_{\dot \alpha}  X_{ef} 
 \bar D^{\dot \alpha}  X_{gh} 
\big]
\eeq
with arbitrary real functions $g^{ab\cdots}_{\#}$ of 
the $G$-invariants $X_{mn}$. 
From this equation,
the components of $\Lambda_{ik\bar j\bar l}$ can be read as
\beq
&&
\Lambda_{ik\bar j\bar l} =
\Big[
g_1^{ab} (X_{mn})  
(v_a  X_J^\dagger  X_L^\dagger \xi^\dagger\xi X_K X_I v_b) 
\non  
&& \hs{12} 
+g_2^{abcd} (X_{mn}) 
      (v_a X_J^\dagger \xi^\dagger\xi  X_I v_b)  
      (v_c X_L^\dagger \xi^\dagger\xi  X_K v_d)  
\non
&& \hs{12} 
   +
 g_3^{abcd} (X_{mn}) 
   (v_a \xi^\dagger\xi X_K X_I v_b)  
   (v_c  X_J^\dagger  X_L^\dagger \xi^\dagger\xi v_d)     
\non && \hs{12} 
+ g_4^{abcdef}  (X_{mn})
(v_a \xi^\dagger\xi  X_I v_b)  
(v_c \xi^\dagger\xi  X_K v_d)  
(v_e X_J^\dagger X_L^\dagger \xi^\dagger\xi  v_f)  
\non && \hs{12} 
+ g_5^{abcdef}  (X_{mn})
(v_a \xi^\dagger\xi  X_K  X_I v_b)   
(v_c X_J^\dagger \xi^\dagger\xi  v_d)  
(v_e X_L^\dagger \xi^\dagger\xi  v_f)  
\non && \hs{12} 
+ g_6^{abcdef}  (X_{mn})
(v_a \xi^\dagger\xi  X_I v_b)  
(v_c X_J^\dagger \xi^\dagger\xi  X_K v_d)  
(v_e X_L^\dagger \xi^\dagger\xi  v_f)  
\non && \hs{12} 
+ g_7^{abcdefgh}  (X_{mn})
(v_a \xi^\dagger\xi  X_I v_b)  
(v_c \xi^\dagger\xi  X_K v_d)  
(v_e X_J^\dagger \xi^\dagger\xi  v_f)  
(v_g X_L^\dagger \xi^\dagger\xi  v_h)  
\Big]
\non
&&
\hs{12}
\times  E^I_i (\Phi)   E^K_k (\Phi) 
  E^{*J}_j (\Phi^\dagger) E^{*L}_l (\Phi^\dagger) .
  \label{eq:A-type-4deriv}
\eeq
Note that Eq.~(\ref{eq:A-type-4deriv}) contains 
the multiple functions labeled by $ab\cdots$, 
implying more general than Eq.~(\ref{eq:chiral-4deriv-A}).
 
Derivative terms higher than four derivatives 
can be constructed by using space-time derivative on $X_{ab}$.
For instance, six-derivative terms can be constructed as
\beq
&&
\mathcal{L}^{(6)} =  \frac{1}{16} \int \! d^4 \theta \ 
\sum_{p=1,2}
Y_p 
\big[
h_{1,p}^{ab} (X_{mn}) 
 \bar D_{\dot \alpha} \bar D^{\dot \alpha} 
 D^{\alpha} D_{\alpha} X_{ab} 
\non && \hs{20}
+ 
h_{2,p}^{abcd} (X_{mn})
 \bar D_{\dot \alpha}  D^{\alpha} X_{ab}  
 \bar D^{\dot \alpha}  D_{\alpha} X_{cd}  
+  
h_{3,p}^{abcd} (X_{mn})  
 D^{\alpha}   D_{\alpha} X_{ab}  
 \bar D_{\dot \alpha}  \bar D^{\dot \alpha} X_{cd}
\non &&  \hs{20}
+  
h_{4,p}^{abcdef} (X_{mn})  
 D^{\alpha} X_{ab}  D_{\alpha} X_{cd}  
 \bar D_{\dot \alpha}  \bar D^{\dot \alpha} X_{ef}
+  
h_{5,p}^{abcdef} (X_{mn})  
 D^{\alpha}  D_{\alpha}  X_{ab}    
 \bar D_{\dot \alpha}  X_{cd}   \bar D^{\dot \alpha} X_{ef}
\non && \hs{20}
+ h_{6,p}^{abcdef}  (X_{mn}) 
  D^{\alpha} X_{ab} D_{\alpha} \bar D_{\dot \alpha}  X_{cd} 
 \bar D^{\dot \alpha}  X_{ef} 
\big]
\eeq
with arbitrary functions $h_{\#,p}^{ab\cdots}$ of of the $G$-invariants 
$X_{mn}$ 
and the extra derivative terms $Y_p$ ($p=1,2$) defined by
\beq
Y_1 = \del_m \del^m X_{a'b'} 
, \quad 
Y_2 =\del_m X_{a'b'}   \del^m X_{c'd'}  .
\eeq

%%%%%%%%%%%%%%%%%%%%%%%%%%%%%%%%
\subsection{C-type}\label{sec:hd-C-type}

Here, we discuss the construction of higher 
derivative terms from the C-type invariants. 
In the geometrical method, 
$G$-invariant four-derivative terms are given by
\beq
&& \mathcal{L}^{(4)} 
=  \frac{1}{16} \int \! d^4 \theta \ 
\Lambda_{ik\bar{j} \bar{l}} (\Phi, \Phi^{\dagger}) 
 D^{\alpha} \Phi^i
 D_{\alpha} \Phi^k \bar{D}_{\dot{\alpha}} \Phi^{\dagger \bar j}
 \bar{D}^{\dot{\alpha}} \Phi^{\dagger \bar{l}} 
, \non 
&&  
\Lambda_{ik\bar j\bar l} 
= 
w_1 (\tr (P_aP_b),\cdots) g_{(i\bar j} g_{k\bar l)} 
 + w_2 (\tr (P_aP_b),\cdots) R_{i\bar jk\bar l} 
\non &&\hs{10}
 + w_3 (\tr (P_aP_b),\cdots ) R_{(i\bar j} R_{k\bar l)}
+ w_4 (\tr (P_aP_b),\cdots) g_{(i\bar{j}} R_{k \bar{l})}.
  \label{eq:chiral-4deriv-C}
\eeq
As the A-type case, 
$w_{1,2,3,4}$  are arbitrary functions 
of the strict $G$-invariants $\tr (P_aP_b)$, 
$\tr (P_aP_bP_c)$ and so on.

In the group theoretical method, 
four-derivative terms can be constructed 
from the C-type projectors $P_a$ 
and the supercovariant derivatives
$D_{\alpha}$ and $\bar D^{\dot\alpha}$. 
All possible $G$-invariant terms 
$X_A(D,\bar D; P_a,P_b,\cdots)$ 
including $P_a$ and two $D$'s and two $\bar D$'s 
are summarized in Table \ref{table:c-type}.
These terms are classified by 
the number of traces and the number of $P_a$,
where each trace should contain more than two $P_a$'s 
with different $a ,b, c \cdots$. 
Then, the four-derivative term constructed from 
the C-type can be written as
\beq
\mathcal{L}^{(4)} 
=  \frac{1}{16} \int \! d^4 \theta \ 
\sum_{A;a,b,\cdots} 
g^A_{ab\cdots} 
(\tr (P_c P_d),\cdots)
 X_A(D,\bar D; P_a,P_b,\cdots)\;\;
\eeq
where 
$X_A(D,\bar D; P_a,P_b,\cdots)$ are 
the $G$-invariant four-derivative terms
given in Table \ref{table:c-type} 
and 
$g^A_{ab\cdots}$ are {\it arbitrary} functions 
of the C-type $G$-invariants $\tr (P_c P_d)$, 
$\tr (P_c P_d P_e)$ and so on.

%%%%%%%%%%%%%%%
\begin{table}
\begin{center}
\begin{tabular}{c||c|c|c}
 $\# P$ $\backslash$  $\#$ tr & 1 & 2 & 3 \\ \hline\hline
2 & $\tr (D^{\alpha} D_{\alpha} P_a 
 \bar D_{\dot \alpha}  \bar D^{\dot \alpha} P_b)$ 
   &  non & non \\\hline
3 & $\tr (P_a D^{\alpha} D_{\alpha} P_b 
 \bar D_{\dot \alpha}  \bar D^{\dot \alpha} P_c)$ & non & non\\ \hline
4 & $\tr (D^{\alpha} P_a D_{\alpha} P_b 
  \bar D_{\dot \alpha}  P_c \bar D^{\dot \alpha} P_d)$ & 
 $\tr (D^{\alpha} P_a D_{\alpha} P_b) 
  \tr (\bar D_{\dot \alpha}  P_c \bar D^{\dot \alpha} P_d)$
 & non \\
 &
$\tr ( P_a D^{\alpha} D_{\alpha} P_b \cdot 
   P_c \bar D_{\dot \alpha}  \bar  D^{\dot \alpha} P_d)$ & 
 $\tr (D^{\alpha} P_a \bar D_{\dot \alpha} P_b) 
  \tr (D_{\alpha}  P_c \bar D^{\dot \alpha} P_d)$
 & \\\hline
5 & $\cdots$ && $\cdots$
\end{tabular}
\caption{Four-derivative terms 
$X_A(D,\bar D; P_a,P_b,\cdots)$ constructed 
from the C-type invariants.
The columns denote the number of traces, 
and the lows denote the number of $P_a$.
Each trace contains more than two $P_a$'s 
with different $a ,b, c \cdots$.}
\label{table:c-type}
\end{center}
\end{table}
%%%%%%%%%%%%

This method can be generalized to 
 derivative terms higher than four derivatives.
It can be achieved by
allowing 
$g^A_{ab\cdots}$ to contain linear terms 
including space-time derivatives or allowing 
$X_A$ to contain space-time derivatives. 
For instance,  six-derivative terms can be constructed as
\beq
&& 
\mathcal{L}^{(6)} =  \frac{1}{16} \int \! d^4 \theta \ 
\Big[
\sum_{p=1,2; A; a,b,\cdots } 
 h_{A,p}^{ab\cdots} (\tr (P_e P_f),\cdots) \tr (Y_p) X_A(D,\bar D; P_a,P_b,\cdots)
\non
&& \hs{20} 
+\sum_{p=1,2; a,b,\cdots}
\big\{
H_{1,p}^{ab\cdots} (\tr (P_e P_f),\cdots)
\tr (Y_p D^{\alpha} D_{\alpha} P_a 
 \bar D_{\dot \alpha}  \bar D^{\dot \alpha} P_b)
\non && \hs{30} 
+
H_{2,p}^{ab\cdots} (\tr (P_e P_f),\cdots)
\tr (Y_p P_a D^{\alpha} D_{\alpha} P_b 
 \bar D_{\dot \alpha}  \bar D^{\dot \alpha} P_c) 
\non && \hs{30} 
+ H_{3,p}^{ab\cdots} (\tr (P_e P_f),\cdots)
  \tr (Y_p D^{\alpha} P_a D_{\alpha} P_b 
  \bar D_{\dot \alpha}  P_c \bar D^{\dot \alpha} P_d)
\non && \hs{30} 
+H_{4,p}^{ab\cdots} (\tr (P_e P_f),\cdots)
 \tr ( Y_p P_a D^{\alpha} D_{\alpha} P_b \cdot 
   P_c \bar D_{\dot \alpha}  \bar  D^{\dot \alpha} P_d)
\big\} 
+ \cdots
\Big]
\label{6-derivative}
\eeq
with arbitrary functions 
$h_{A,p}^{ab\cdots}$ and $H_{A,p}^{ab\cdots}$ 
of the C-type $G$-invariants, and 
the extra two-derivative terms $Y_p$ ($p=1,2$) given  by
\beq
 Y_1 =\del_m \del^m P_{a'} , \quad 
 Y_2 = \del_m P_{a'} \del^m P_{b'}.
\eeq
The dots in Eq.~\eqref{6-derivative} imply multi-trace terms such as
$\tr(\partial_m P_{a^\prime}D^\alpha D_\alpha P_a)
  \tr(\partial^m P_{b^\prime}\bar D_{\dot\alpha}\bar D^{\dot\alpha}P_b)$
 and so on.

%%%%%%%%%%%%%%%%%%%%%%
\section{Supersymmetric Chiral Symmetry Breaking}
\label{sec:chisb}

In this section, we 
show an explicit example of higher derivative interactions of quasi-NG bosons.
We consider higher derivative corrections for 
supersymmetric chiral symmetry breaking, which is a maximal realization 
with each massless chiral superfield containing one 
 NG boson and one quasi-NG boson.

\subsection{Supersymmetric chiral Lagrangian}

Let us consider the chiral symmetry breaking
\begin{align}
 G = SU(N)_{\rm L} \times SU(N)_{\rm R} 
\to H = SU(N)_{\rm L+R}.
\end{align}
The corresponding NG modes 
span the coset space
\beq
 G/H 
= \frac{SU(N)_{\rm L} \times SU(N)_{\rm R}}{SU(N)_{\rm L+R}}
\simeq 
SU(N).
\eeq
We denote generators of the coset by
$T_A \in {\cal SU}(N)$.
It was shown in Ref.~\cite{Lerche:1983qa} that 
a vacuum expectation value belonging to 
a real representation 
gives rise to the same numbers of quasi-NG bosons 
and NG bosons, 
which  is a maximal realization.
The chiral symmetry breaking belongs to this class, 
and the total target space is
\beq 
 \GC/\hat H \simeq SU(N)^{\mathbb C} 
= \GC/\HC \simeq 
 SL(N,{\mathbb C}) \simeq T^* SU(N).
\eeq
The coset representative is written as
\begin{align}
 M = \exp (i \Phi^A T_A) 
\in \GC/\hat H, 
\label{pion_superfield}
\end{align}
where the NG superfields
are in the form of 
\begin{align}
\Phi^A (y,\theta) 
= \pi^A (y) + i \sigma^A (y) 
+ \theta \psi^A (y) + \theta\theta F^A(y), 
\end{align}
with NG bosons $\pi^A$,  
quasi-NG bosons $\sigma^A$, 
and quasi-NG fermions $\psi^A$.

The nonlinear transformation law of 
the NG supermultiplets is 
\beq
 M \to M' = g_{\rm L} M g_{\rm R} , \quad
 (g_{\rm L}, g_{\rm R}) 
\in SU(N)_{\rm L} \times SU(N)_{\rm R} .
\eeq
From the transformation
\beq
  M M^\dagger \to g_{\rm L}  M M^\dagger g_{\rm L}^\dagger,
\eeq
the simplest K\"ahler potential 
is found to be 
\beq
 K_0 =  f_{\pi}^2  \tr (M M^\dagger),\label{eq:chiral-Kahler0}
\eeq
where $f_{\pi}$ is a constant.
Therefore, the leading order of the 
bosonic part of the Lagrangian 
in the derivative expansion reads  
\beq
 {\cal L}_0 = - f_{\pi}^2 \tr (\del_m M \del^m M^\dagger),
\eeq
where $M$ is the lowest component of the NG superfield \eqref{pion_superfield}.
However, the K\"ahler potential in Eq.~(\ref{eq:chiral-Kahler0}) is not general. 
In fact, the most general K\"ahler potential can be written as 
\cite{Kotcheff:1988ji,Nitta:1998qp}
\beq
 K = f(\tr (M M^\dagger),  \tr [(M M^\dagger)^2], \cdots, 
\tr [(M M^\dagger)^{N-1}] )  \label{eq:chiral-Kahler-general}
\eeq
with an {\it arbitrary} function of $N-1$ variables. 
The physical reason why we have 
an arbitrary function is the existence of the quasi-NG bosons.
Since the isometry of the target manifold is $G$ 
but not $\GC$, the target manifold is not homogeneous.
One can deform the shape of the target manifold 
along the directions of 
the quasi-NG bosons, with keeping the isometry $G$.\footnote{
If one requires the Ricci-flat condition on the target manifold, 
the arbitrary function is fixed. 
That is known as the Stenzel metric. This is not the scope of this paper.
} 

If we set all quasi-NG bosons to be zero \cite{Shore:1988mn,Kotcheff:1988ji}
\beq
 U = M|_{\sigma^A=0} \in SU(N),\label{eq:QNGzero}
\eeq
we have usual chiral Lagrangian
\beq
 {\cal L} = - f_\pi^2 \tr (\del_m U \del^m  U^\dagger)
= - f_\pi^2 \tr (U^\dagger \del_m U)^2
\eeq
with 
the decay constant $f_\pi$ determined from $f$.

One interesting feature of chiral symmetry breaking 
in supersymmetric vacua is 
that the unbroken group $H = SU(N)_{\rm L+R}$ can be 
further broken to its subgroup 
due to the vacuum expectation value of 
the quasi-NG bosons \cite{Kotcheff:1988ji}.
Some of quasi-NG bosons change to NG bosons 
at less symmetric vacua 
\cite{Kotcheff:1988ji,Nitta:1998qp}

%%%%%%%%%%%%%%%%%%%%%%
\subsection{Higher derivative terms: supersymmetric chiral perturbation}\label{sec:super-chiral-pert}

Let us discuss possible higher derivative terms 
for the supersymmetric chiral Lagrangian.
The simplest candidate of a four-derivative term is
\begin{align}
\mathcal{L}^{(4)}_0 = & \frac{1}{16} \int \! d^4 \theta \ 
\Lambda_{ik\bar{j} \bar{l}}
(\Phi,
 \Phi^{\dagger}) 
D^{\alpha} \Phi^i
 D_{\alpha} \Phi^k \bar{D}_{\dot{\alpha}} \Phi^{\dagger \bar j}
 \bar{D}^{\dot{\alpha}} \Phi^{\dagger \bar{l}} 
= \int \! d^4 \theta \ 
\tr (D^{\alpha} M \bar{D}_{\dot{\alpha}} M^\dagger 
 D_{\alpha} M \bar{D}^{\dot{\alpha}} M^\dagger ),
   \label{eq:chiral-4deriv0}
\end{align}
where components of {
$\Lambda_{ik\bar{j} \bar{l}}$
are determined from the right hand side. 
The bosonic part of this term is
\beq
 \mathcal{L}^{(4)}_{0,b} = \tr (\del^m M \del^n M^\dagger \del_m M \del_n M^\dagger)
\eeq
in the canonical branch with $F^A=0$.

However, Eq.~(\ref{eq:chiral-4deriv0}) is not general.
As in the leading term, we have a freedom 
to deform the tensor along the directions of the quasi-NG bosons.
The most general Lagrangian can be written as
\beq
&& \mathcal{L}^{(4)} =  \frac{1}{16} \int \! d^4 \theta \ 
\Lambda_{ik\bar{j} \bar{l}}
(\Phi,
 \Phi^{\dagger}) 
D^{\alpha} \Phi^i
 D_{\alpha} \Phi^k \bar{D}_{\dot{\alpha}} \Phi^{\dagger \bar j}
 \bar{D}^{\dot{\alpha}} \Phi^{\dagger \bar{l}} \non
&&\hs{10}
= \int \! d^4 \theta \ 
\Big[ \sum_{k=1}^{N-1} 
g_1^k(\tr (M M^\dagger),  \cdots, \tr [(M M^\dagger)^{N-1}] )
\tr (D^{\alpha} M \bar{D}_{\dot{\alpha}} M^\dagger 
 D_{\alpha} M \bar{D}^{\dot{\alpha}} M^\dagger (MM^\dagger)^k) 
\non && \hs{25}
+
\sum_{k,l=1}^{N-1} 
 g_2^{kl}(\tr (M M^\dagger),  \cdots, \tr [(M M^\dagger)^{N-1}] ) 
\non && \hs{30}
\times
\tr (D^{\alpha} M \bar{D}_{\dot{\alpha}} M^\dagger (MM^\dagger)^k ) 
\tr (D_{\alpha} M \bar{D}^{\dot{\alpha}} M^\dagger  (MM^\dagger)^l)\Big] 
   \label{eq:chiral-4deriv}
\eeq
with an arbitrary functions $g_{1}^k$ and $g_{2}^{kl}$ of 
$N-1$ $G$-invariants $\tr (MM^\dagger),\cdots , \tr
(MM^\dagger)^{N-1}$. 
The bosonic part of this term is
\beq
&& \mathcal{L}^{(4)}_b = 
\sum_{k=1}^{N-1} 
g_1^k (\tr (M M^\dagger),   \cdots, 
\tr [(M M^\dagger)^{N-1}] ) \,
\tr (\del^m M \del^n M^\dagger \del_m M \del_n M^\dagger  (MM^\dagger)^k) 
\non
&&
+ 
\sum_{k,l=1}^{N-1} 
g_2^{kl}(\tr (M M^\dagger),  \cdots, 
\tr [(M M^\dagger)^{N-1}] ) \,
\tr (\del^m M \del^n M^\dagger (MM^\dagger)^k )
\tr(\del_m M \del_n M^\dagger (MM^\dagger)^l ).
\non
\eeq
If we set all quasi-NG bosons to be zero as in Eq.~(\ref{eq:QNGzero}), 
\beq  
 M M^\dagger|_{\sigma^A =0} = 
 U U^\dagger = {\bf 1}_{N} \quad
 (\tr [(M M^\dagger)^k]|_{\sigma^A =0} =N),
\eeq
 and 
the bosonic part in the canonical branch with $F^A=0$  
becomes
\beq
 \mathcal{L}^{(4)}_{b}|_{\sigma=0} 
= g_{1,0} \tr (\del^m U \del^n U^\dagger \del_m U \del_n U^\dagger) 
 +g_{2,0} \tr (\del^m U \del^n U^\dagger ) 
          \tr(\del_m U \del_n U^\dagger)
\eeq
with $g_{1,0} = \sum_{k=1}^{N-1} g_1^k (N,\cdots,N)$ 
and $g_{2,0} = \sum_{k,l=1}^{N-1} g_2^{k,l} (N,\cdots,N)$. 
One notes that the term 
$\tr (\del^m U \del_m U^\dagger \del^n U \del_n U^\dagger)$
or $\tr (\del^m U \del_m U^\dagger ) 
          \tr(\del^n U \del_n U^\dagger)$ 
is not allowed as a bosonic part of the 
supersymmetric Lagrangian.

Next, let us construct six-derivative terms. 
They can be written as
\beq
&& \mathcal{L}^{(6)} 
= \int \! d^4 \theta \ \Big[
\sum_{k=1}^{N-1}
h_1^k(\tr (M M^\dagger), \cdots 
) 
\tr (
\del_m M \del^m M^\dagger
D^{\alpha} M \bar{D}_{\dot{\alpha}} M^\dagger 
 D_{\alpha} M \bar{D}^{\dot{\alpha}} M^\dagger 
(M M^\dagger)^k
)
\non && \hs{20}
+
\sum_{k,l=1}^{N-1}
h_2^{kl}(\tr (M M^\dagger), \cdots 
) 
\tr (
\del_m M \del^m M^\dagger
D^{\alpha} M \bar{D}_{\dot{\alpha}} M^\dagger 
 (M M^\dagger)^k)
\non && \hs{30}
\times \tr(
 D_{\alpha} M \bar{D}^{\dot{\alpha}} M^\dagger 
(M M^\dagger)^l
)
\non &&  \hs{20}
+ 
\sum_{k,l,j=1}^{N-1}
h_3^{klj}(\tr (M M^\dagger), \cdots 
) 
\tr (\del_m M \del^m M^\dagger (M M^\dagger)^k)
\non
&&\hs{30}
\times
\tr (D^{\alpha} M \bar{D}_{\dot{\alpha}} M^\dagger (M M^\dagger)^l) 
\tr (D_{\alpha} M \bar{D}^{\dot{\alpha}} M^\dagger (M M^\dagger)^j)
 \Big]
\label{eq:chiral-6deriv}
\eeq
with arbitrary functions $h_1^k,h_2^{kl},h_3^{klj}$ of 
$N-1$ $G$-invariants 
$\tr (MM^\dagger),\cdots , \tr (MM^\dagger)^{N-1}$. 
The dots in Eq.~\eqref{eq:chiral-6deriv} imply multi-trace terms such as
\begin{align}
  \tr(D^{\alpha} M  \partial^m M^\dagger
    D_{\alpha} M \bar{D}_{\dot{\alpha}} M^\dagger (M M^\dagger)^k)
  \tr(\partial_m M \bar{D}^{\dot{\alpha}} M^\dagger (M M^\dagger)^l)
\notag 
\end{align}
and 
\begin{align}
\tr(\partial_m M  \bar{D}_{\dot{\alpha}} M^\dagger
    D^{\alpha} M \bar{D}^{\dot{\alpha}} M^\dagger (M M^\dagger)^k)
   \tr( D_{\alpha} M \partial^m M^\dagger (M M^\dagger)^l).
\notag 
\end{align}
The bosonic part of this term is
\beq
&& 
\mathcal{L}^{(6)}_b 
= 
\sum_{k=1}^{N-1} \Big[
h_1^k(\tr (M M^\dagger), \cdots 
) 
\tr (
 \del_m M \del^m M^\dagger
 \del^n M \del^o M^\dagger 
 \del_n M \del_o M^\dagger 
(M M^\dagger)^k
)
\non && \hs{7}
+
\sum_{k,l=1}^{N-1}
h_2^{kl}(\tr (M M^\dagger), \cdots 
) 
\tr (
 \del_m M \del^m M^\dagger
 \del^n M \del_o M^\dagger 
  (M M^\dagger)^k)
\tr( 
 \del_n M \del^o M^\dagger 
  (M M^\dagger)^l
)
\non && \hs{7}
+ 
\sum_{k,l,j=1}^{N-1}
h_3^{klj}(\tr (M M^\dagger), \cdots 
) 
\tr (\del_m M \del^m M^\dagger (M M^\dagger)^k)
\non && \hs{18}
\times
\tr (\del^n M \del_o M^\dagger (M M^\dagger)^l) 
\tr (\del_n M \del^o M^\dagger (M M^\dagger)^j)
 \Big] 
+ \cdots.
\label{eq:chiral-6deriv-boson}
\eeq
If we set all quasi-NG bosons to be zero, these terms reduce to
\beq
&& 
\mathcal{L}^{(6)}_b|_{\sigma=0} 
= 
h_{1,0}  \,
\tr (
 \del_m U \del^m U^\dagger
 \del^n U \del^o U^\dagger 
 \del_n U \del_o U^\dagger )
\non
&& \hs{15}
+ h_{2,0} \,
\tr ( \del_m U \del^m U^\dagger
       \del^n U \del_o U^\dagger )
\tr ( \del_n U \del^o U^\dagger ) 
\non && \hs{15}
+ h_{3,0} \,
\tr (\del_m U \del^m U^\dagger )
\tr (\del^n U \del_o U^\dagger ) 
\tr (\del_n U \del^o U^\dagger )
+ \cdots.
\eeq
with $h_{1,0} = \sum_{k=1}^{N-1} h_1^k (N,\cdots,N)$, 
$h_{2,0} = \sum_{k,l=1}^{N-1} h_2^{k,l} (N,\cdots,N)$ and 
$h_{3,0} = \sum_{k,l,j=1}^{N-1} h_3^{k,l,j} (N,\cdots,N)$.  
We can construct the eight- or higher 
derivative terms in the same way.

%%%%%%%%%%%%%%%%%%%%%%
\section{Conclusion and discussions} \label{sec:conc}

In this paper we have constructed higher derivative correction terms for
massless NG and quasi-NG bosons and fermions in the manifestly supersymmetric
off-shell formalism. 
In general, when a global symmetry is broken in supersymmetric vacua, massless
quasi-NG bosons and fermions appear. Low-energy effective theories are
governed by supersymmetric nonlinear sigma models of the NG and quasi-NG fields.
The number of the quasi-NG fields is determined by the
structure of the coset group $G^{\mathbb{C}}/\hat{H}$.
The $G$-invariant 
K\"ahler potentials of the nonlinear sigma models are classified into
A-,B-, and C-types. We have shown the $G$-invariant quantities and
examples of K\"ahler potentials.

In superfield formalism,
the higher derivative term in the chiral model is 
given by 
a (2,2) K\"ahler tensor $\Lambda_{ij\bar k \bar l}$ 
symmetric in holomorphic and anti-holomorphic indices, 
whose components are functions of the chiral superfields
$\Phi^i$. 
By using this formalism  
we have constructed higher derivative corrections 
to supersymmetric nonlinear realizations. 
The tensors $\Lambda_{ij\bar{k} \bar{l}}$ are constructed by
the $G$-invariant K\"ahler metrics in the A-,B-,C-types.
Remarkably, in the A-,C-types, the tensors $\Lambda_{ij\bar{k}\bar{l}}$
include degrees of freedom for the strict $G$-invariant quantities
$X_{ab}$ and $\mathrm{tr} (P_a P_b \cdots)$. 
For the B-type, this is the pure realization, and there are no quasi-NG modes. 
We have found 
that the higher derivative terms are unique 
up to constants.
For the A- and C-types, there are quasi-NG modes and 
higher derivative terms contain 
arbitrary functions
which depends on the strict $G$-invariants.
We have also constructed the higher derivative terms in purely group
theoretical manners.
As a practical example,  
we have further studied the case of 
chiral symmetry breaking in more detail.

Several discussions are addressed here.

In this paper, we have studied spontaneous breaking of 
exact symmetry leading 
to exactly massless NG bosons and quasi-NG bosons (fermions).
For approximate symmetry, an explicit breaking term 
should be introduced 
which give NG bosons masses. 
Consequently, they become pseudo NG bosons, 
such as pions for the chiral symmetry breaking.
In supersymmetric theory, a symmetry breaking potential term can be introduced 
by the superpotential $W$. 
The introduction of the superpotential can be 
treated perturbatively, which was done at least 
for single component cases \cite{SaYaYo,Nitta:2014pwa}.

As for another future work, the inclusion of the supersymmetric 
WZW term 
\cite{Gates:2000rp,Banin:2006db} 
should be discussed for 
supersymmetric chiral perturbation theory. 
For supersymmetric 
chiral perturbation theory with general target spaces, 
K\"ahler normal coordinates \cite{Higashijima:2000wz} 
should be useful 
as in Ref.~\cite{Banin:2006db}.

A BPS Skyrme model was discovered some years back \cite{Adam:2010fg},
which consists of only the sixth-order higher derivative term 
as well as appropriate potentials. 
Our result should be useful to investigate 
supersymmetric version of this model.

In this paper, we have considered the canonical branch 
with $F=0$ for solutions to the auxiliary field equations. 
It is known 
for the ${\mathbb C}P^1$ model that 
there is also a non-canonical branch 
with $F\neq 0$ \cite{AdQuSaGuWe,Nitta:2014pwa}.
While the usual kinetic term disappears  in this case, 
the theory admits a baby Skyrmion \cite{AdQuSaGuWe},  
which was shown to be a 1/4 BPS state \cite{Nitta:2014pwa}.
Investigating 
non-canonical branches and 1/4 BPS baby Skyrmions 
for general K\"ahler $G/H$ 
 are one of interesting future directions.

The supersymmetric 
${\mathbb C}P^{N-1}$ model with four supercharges also appears as
the world-volume effective action of a BPS non-Abelian vortex 
in ${\cal N}=2$ supersymmetric $U(N)$ gauge theory with 
$N$ hypermultiplets in the fundamental representation \cite{Hanany:2003hp}. 
Higher derivative corrections to 
the effective action were calculated in 
Ref.~\cite{Eto:2012qda}.
It was shown that 
1/2 BPS lumps (sigma model instantons) 
are not modified in the presence of higher derivative terms  
\cite{Nitta:2014pwa,Eto:2012qda}.
This should be so because 
a composite state of lumps inside a non-Abelian vortex 
is a 1/4 BPS state and it is nothing but a Yang-Mills 
instanton in the bulk point of view 
\cite{Eto:2004rz}. 
See Refs.~\cite{Tong:2005un,Eto:2006pg,Shifman:2007ce} 
for a review of BPS composite solitons.

As this regards, 
some other K\"ahler $G/H$ manifolds are realized 
on a vortex in supersymmetric gauge theories with 
gauge groups $G$ \cite{Eto:2008yi}.  
In particular, the cases of $G=SO(N),USp(N)$
 were studied in detail \cite{Eto:2009bg}.
Therefore, 1/2 BPS lumps in sigma models on 
K\"ahler $G/H$ 
with higher derivative terms 
describe instantons in gauge theories with gauge group $G$.
It should be checked whether 
higher derivative corrections for lumps in these cases  
are canceled out.

The supersymmetric chiral Lagrangian 
studied in Sec.~\ref{sec:chisb} also appears 
as the effective theory on BPS non-Abelian domain walls 
in ${\cal N}=2$ supersymmetric $U(N)$ gauge theories 
with $2N$ hypermultiplets in the fundamental representation 
with mass $\pm m$ \cite{Shifman:2003uh}.
A four-derivative correction was 
partly derived in Ref.~\cite{Eto:2005cc}.  

A general framework of a superfield formulation of the effective theories  on BPS soliton world-volumes 
with four supercharges 
was formulated in Ref.~\cite{Eto:2006uw}. 
This should be generalized to the case with 
higher derivative corrections.

%%%%%%%%%%%%%%
\subsection*{Acknowledgments}
The work of M.\ N.\ is supported in part by Grant-in-Aid for 
Scientific Research (No. 25400268) and by the ``Topological 
Quantum Phenomena''  Grant-in-Aid for Scientific Research on 
Innovative Areas (No. 25103720) from the Ministry of Education, 
Culture, Sports, Science and Technology  (MEXT) of Japan.
The work of S.~S. is supported in part by Kitasato University Research Grant for Young
Researchers.

%%%%%%%%%%%%%%%%%%%%%%%%%%%%%%%%%%%%%%%%%%%%%%%%%


\begin{thebibliography}{99}

%\cite{Coleman:1969sm}
\bibitem{Coleman:1969sm} 
  S.~R.~Coleman, J.~Wess and B.~Zumino,
  ``Structure of phenomenological Lagrangians. 1.,''
  Phys.\ Rev.\  {\bf 177}, 2239 (1969);
  %%CITATION = PHRVA,177,2239;%%
  %1409 citations counted in INSPIRE as of 29 Jul 2014
%\cite{Callan:1969sn}
%\bibitem{Callan:1969sn} 
  C.~G.~Callan, Jr., S.~R.~Coleman, J.~Wess and B.~Zumino,
  ``Structure of phenomenological Lagrangians. 2.,''
  Phys.\ Rev.\  {\bf 177}, 2247 (1969).
  %%CITATION = PHRVA,177,2247;%%
  %1217 citations counted in INSPIRE as of 29 Jul 2014


\bibitem{Leutwyler:1993iq} 
  H.~Leutwyler,
  ``On the foundations of chiral perturbation theory,''
  Annals Phys.\  {\bf 235}, 165 (1994)
  [hep-ph/9311274].
  %%CITATION = HEP-PH/9311274;%%
  %281 citations counted in INSPIRE as of 21 Jun 2014

\bibitem{Seiberg:1994rs} 
  N.~Seiberg and E.~Witten,
  ``Electric - magnetic duality, monopole condensation, and confinement in N=2 supersymmetric Yang-Mills theory,''
  Nucl.\ Phys.\ B {\bf 426}, 19 (1994)
  [Erratum-ibid.\ B {\bf 430}, 485 (1994)]
  [hep-th/9407087],
  %%CITATION = HEP-TH/9407087;%%
  %2509 citations counted in INSPIRE as of 25 Jun 2014
%  N.~Seiberg and E.~Witten,
  ``Monopoles, duality and chiral symmetry breaking in N=2 supersymmetric QCD,''
  Nucl.\ Phys.\ B {\bf 431}, 484 (1994)
  [hep-th/9408099].
  %%CITATION = HEP-TH/9408099;%%
  %1831 citations counted in INSPIRE as of 25 Jun 2014

%\cite{Buchmuller:1982xn}
\bibitem{Buchmuller:1982xn} 
  W.~Buchmuller, S.~T.~Love, R.~D.~Peccei and T.~Yanagida,
  ``Quasi Goldstone Fermions,''
  Phys.\ Lett.\ B {\bf 115}, 233 (1982).
  %%CITATION = PHLTA,B115,233;%%
  %125 citations counted in INSPIRE as of 01 Aug 2014
%\cite{Buchmuller:1983iu}
%\bibitem{Buchmuller:1983iu} 
  W.~Buchmuller, R.~D.~Peccei and T.~Yanagida,
  ``Quasi Nambu-Goldstone Fermions,''
  Nucl.\ Phys.\ B {\bf 227}, 503 (1983).
  %%CITATION = NUPHA,B227,503;%%
  %134 citations counted in INSPIRE as of 01 Aug 2014




%\cite{Buchmuller:1982tf}
\bibitem{Buchmuller:1982tf} 
  W.~Buchmuller, R.~D.~Peccei and T.~Yanagida,
  ``Quarks and Leptons as Quasi Nambu-Goldstone Fermions,''
  Phys.\ Lett.\ B {\bf 124}, 67 (1983);
  %%CITATION = PHLTA,B124,67;%%
  %191 citations counted in INSPIRE as of 01 Aug 2014
%\cite{Buchmuller:1983na}
%\bibitem{Buchmuller:1983na} 
  W.~Buchmuller, R.~D.~Peccei and T.~Yanagida,
  ``Weak Interactions of Quasi Nambu-goldstone Fermions,''
  Nucl.\ Phys.\ B {\bf 231}, 53 (1984).
  %%CITATION = NUPHA,B231,53;%%
  %73 citations counted in INSPIRE as of 01 Aug 2014


%\cite{Zumino:1979et}
\bibitem{Zumino:1979et} 
  B.~Zumino,
  ``Supersymmetry and Kahler Manifolds,''
  Phys.\ Lett.\ B {\bf 87}, 203 (1979).
  %%CITATION = PHLTA,B87,203;%%
  %553 citations counted in INSPIRE as of 02 Aug 2014

%\cite{Kugo:1983ma}
\bibitem{Kugo:1983ma} 
  T.~Kugo, I.~Ojima and T.~Yanagida,
  ``Superpotential Symmetries and Pseudonambu-goldstone Supermultiplets,''
  Phys.\ Lett.\ B {\bf 135}, 402 (1984).
  %%CITATION = PHLTA,B135,402;%%
  %69 citations counted in INSPIRE as of 29 Jul 2014

%\cite{Bando:1983ab}
\bibitem{Bando:1983ab} 
  M.~Bando, T.~Kuramoto, T.~Maskawa and S.~Uehara,
  ``Structure of Nonlinear Realization in Supersymmetric Theories,''
  Phys.\ Lett.\ B {\bf 138}, 94 (1984);
  %%CITATION = PHLTA,B138,94;%%
  %74 citations counted in INSPIRE as of 29 Jul 2014
%\cite{Bando:1984cc}
%\bibitem{Bando:1984cc} 
  M.~Bando, T.~Kuramoto, T.~Maskawa and S.~Uehara,
  ``Nonlinear Realization in Supersymmetric Theories,''
  Prog.\ Theor.\ Phys.\  {\bf 72}, 313 (1984);
  %%CITATION = PTPKA,72,313;%%
  %68 citations counted in INSPIRE as of 29 Jul 2014
%\cite{Bando:1984fn}
%\bibitem{Bando:1984fn} 
  M.~Bando, T.~Kuramoto, T.~Maskawa and S.~Uehara,
  ``Nonlinear Realization in Supersymmetric Theories. 2.,''
  Prog.\ Theor.\ Phys.\  {\bf 72}, 1207 (1984).
  %%CITATION = PTPKA,72,1207;%%
  %56 citations counted in INSPIRE as of 29 Jul 2014

%\cite{Itoh:1985ha}
\bibitem{Itoh:1985ha} 
  K.~Itoh, T.~Kugo and H.~Kunitomo,
  ``Supersymmetric Nonlinear Realization for Arbitrary Kahlerian Coset Space $G/H$,''
  Nucl.\ Phys.\ B {\bf 263}, 295 (1986);
  %%CITATION = NUPHA,B263,295;%%
  %54 citations counted in INSPIRE as of 29 Jul 2014
%\cite{Itoh:1985jz}
%\bibitem{Itoh:1985jz} 
  K.~Itoh, T.~Kugo and H.~Kunitomo,
  ``Supersymmetric Nonlinear Lagrangians of Kahlerian Coset Spaces $G/H$: $G$ = E6, E7 and E8,''
  Prog.\ Theor.\ Phys.\  {\bf 75}, 386 (1986).
  %%CITATION = PTPKA,75,386;%%
  %47 citations counted in INSPIRE as of 04 Aug 2014


%\cite{Bando:1987br}
\bibitem{Bando:1987br} 
  M.~Bando, T.~Kugo and K.~Yamawaki,
  ``Nonlinear Realization and Hidden Local Symmetries,''
  Phys.\ Rept.\  {\bf 164}, 217 (1988).
  %%CITATION = PRPLC,164,217;%%
  %886 citations counted in INSPIRE as of 29 Jul 2014

\bibitem{KahlerG/H}
%\cite{Aoyama:1979zj}
%\bibitem{Aoyama:1979zj} 
  S.~Aoyama,
  ``The Supersymmetric U($N$,r) $\sigma$ Model and Its 0(2) Extended Supersymmetry,''
  Nuovo Cim.\ A {\bf 57}, 176 (1980);
  %%CITATION = NUCIA,A57,176;%%
  %30 citations counted in INSPIRE as of 13 Aug 2014
%\cite{Yasui:1984yu}
%\bibitem{Yasui:1984yu} 
  Y.~Yasui,
  ``The Kahler Potential of E6 / Spin (10) $\times$ SO(2),''
  Prog.\ Theor.\ Phys.\  {\bf 72}, 877 (1984);
  %%CITATION = PTPKA,72,877;%%
  %9 citations counted in INSPIRE as of 13 Aug 2014
%\cite{Achiman:1984ku}
%\bibitem{Achiman:1984ku} 
  Y.~Achiman, S.~Aoyama and J.~W.~van Holten,
  ``The Nonlinear Supersymmetric $\sigma$ Model on E6 / SO(10) $\times$ U(1),''
  Phys.\ Lett.\ B {\bf 141}, 64 (1984);
  %%CITATION = PHLTA,B141,64;%%
  %38 citations counted in INSPIRE as of 13 Aug 2014
%\cite{Achiman:1984fh}
%\bibitem{Achiman:1984fh} 
  Y.~Achiman, S.~Aoyama and J.~W.~van Holten,
  ``Symmetry Breaking in Gauged Supersymmetric $\sigma$ Models,''
  Phys.\ Lett.\ B {\bf 150}, 153 (1985);
  %%CITATION = PHLTA,B150,153;%%
  %12 citations counted in INSPIRE as of 13 Aug 2014
%\cite{Achiman:1985ra}
%\bibitem{Achiman:1985ra} 
  Y.~Achiman, S.~Aoyama and J.~W.~van Holten,
  ``Gauged Supersymmetric $\sigma$ Models and E6 / SO(10) $\times$ U(1),''
  Nucl.\ Phys.\ B {\bf 258}, 179 (1985);
  %%CITATION = NUPHA,B258,179;%%
  %27 citations counted in INSPIRE as of 13 Aug 2014
%\cite{Irie:1983cd}
%\bibitem{Irie:1983cd} 
  S.~Irie and Y.~Yasui,
  ``SUPERSYMMETRIC NONLINEAR sigma MODEL ON E8 / SO(10)$\times$SU(3) $\times$ U(1),''
  Z.\ Phys.\ C {\bf 29}, 123 (1985);
%\cite{Yanagida:1985jc}
%\bibitem{Yanagida:1985jc} 
  T.~Yanagida and Y.~Yasui,
  ``Supersymmetric Nonlinear Sigma Models Based On Exceptional Groups,''
  Nucl.\ Phys.\ B {\bf 269}, 575 (1986);
  %%CITATION = NUPHA,B269,575;%%
  %12 citations counted in INSPIRE as of 13 Aug 2014
  %%CITATION = ZEPYA,C29,123;%%
  %21 citations counted in INSPIRE as of 13 Aug 2014
%\cite{GrootNibbelink:1998tz}
%\bibitem{GrootNibbelink:1998tz} 
  S.~Groot Nibbelink and J.~W.~van Holten,
  ``Matter coupling and anomaly cancellation in supersymmetric sigma models,''
  Phys.\ Lett.\ B {\bf 442}, 185 (1998)
  [hep-th/9808147];
  %%CITATION = HEP-TH/9808147;%%
  %15 citations counted in INSPIRE as of 13 Aug 2014
%\cite{GrootNibbelink:2000hu}
%\bibitem{GrootNibbelink:2000hu} 
  S.~Groot Nibbelink, T.~S.~Nyawelo and J.~W.~van Holten,
  ``Construction and analysis of anomaly free supersymmetric SO(2N) / U(N) sigma models,''
  Nucl.\ Phys.\ B {\bf 594}, 441 (2001)
  [hep-th/0008097].
  %%CITATION = HEP-TH/0008097;%%
  %10 citations counted in INSPIRE as of 13 Aug 2014

%\cite{Kotcheff:1988ji}
\bibitem{Kotcheff:1988ji} 
  A.~C.~W.~Kotcheff and G.~M.~Shore,
  ``Kahler $\sigma$ Models From Supersymmetric Gauge Theories,''
  Int.\ J.\ Mod.\ Phys.\ A {\bf 4}, 4391 (1989).
  %%CITATION = IMPAE,A4,4391;%%
  %18 citations counted in INSPIRE as of 29 Jul 2014


%\cite{Shore:1988mn}
\bibitem{Shore:1988mn} 
  G.~M.~Shore,
  ``Geometry of Supersymmetric $\sigma$ Models,''
  Nucl.\ Phys.\ B {\bf 320}, 202 (1989);
  %%CITATION = NUPHA,B320,202;%%
  %18 citations counted in INSPIRE as of 29 Jul 2014
%\cite{Shore:1989sw}
%\bibitem{Shore:1989sw} 
  G.~M.~Shore,
  ``Geometry of Supersymmetric $\sigma$ Models. 2. Fermions, Connections and Currents,''
  Nucl.\ Phys.\ B {\bf 334}, 172 (1990).
  %%CITATION = NUPHA,B334,172;%%
  %13 citations counted in INSPIRE as of 29 Jul 2014

%\cite{Higashijima:1997ph}
\bibitem{Higashijima:1997ph} 
  K.~Higashijima, M.~Nitta, K.~Ohta and N.~Ohta,
  ``Low-energy theorems in N=1 supersymmetric theory,''
  Prog.\ Theor.\ Phys.\  {\bf 98}, 1165 (1997)
  [hep-th/9706219].
  %%CITATION = HEP-TH/9706219;%%
  %18 citations counted in INSPIRE as of 29 Jul 2014

%\cite{Nitta:1998qp}
\bibitem{Nitta:1998qp} 
  M.~Nitta,
  ``Moduli space of global symmetry in N=1 supersymmetric theories and the quasiNambu-Goldstone bosons,''
  Int.\ J.\ Mod.\ Phys.\ A {\bf 14}, 2397 (1999)
  [hep-th/9805038].
  %%CITATION = HEP-TH/9805038;%%
  %20 citations counted in INSPIRE as of 29 Jul 2014


%\cite{Lerche:1983qa}
\bibitem{Lerche:1983qa} 
  W.~Lerche,
  ``On Goldstone Fields in Supersymmetric Theories,''
  Nucl.\ Phys.\ B {\bf 238}, 582 (1984).
  %%CITATION = NUPHA,B238,582;%%
  %85 citations counted in INSPIRE as of 29 Jul 2014

%\cite{Shore:1984bh}
\bibitem{Shore:1984bh} 
  G.~M.~Shore,
  ``Supersymmetric Higgs Mechanism With Nondoubled Goldstone Bosons,''
  Nucl.\ Phys.\ B {\bf 248}, 123 (1984).
  %%CITATION = NUPHA,B248,123;%%
  %34 citations counted in INSPIRE as of 29 Jul 2014

%\cite{Buchmuller:1986zp}
\bibitem{Buchmuller:1986zp} 
  W.~Buchmuller and W.~Lerche,
  ``Geometry and Anomaly Structure of Supersymmetric $\sigma$ Models,''
  Annals Phys.\  {\bf 175}, 159 (1987).
  %%CITATION = APNYA,175,159;%%
  %47 citations counted in INSPIRE as of 29 Jul 2014


%\cite{Higashijima:1999ki}
\bibitem{Higashijima:1999ki} 
  K.~Higashijima and M.~Nitta,
  ``Supersymmetric nonlinear sigma models as gauge theories,''
  Prog.\ Theor.\ Phys.\  {\bf 103}, 635 (2000)
  [hep-th/9911139].
  %%CITATION = HEP-TH/9911139;%%
  %50 citations counted in INSPIRE as of 29 Jul 2014


\bibitem{Gates:1995fx} 
  S.~J.~Gates, Jr.,
  ``Why auxiliary fields matter: The Strange case of the 4-D, N=1 supersymmetric QCD effective action,''
  Phys.\ Lett.\ B {\bf 365}, 132 (1996)
  [hep-th/9508153], 
  %%CITATION = HEP-TH/9508153;%%
  %36 citations counted in INSPIRE as of 21 Jun 2014
%  S.~J.~Gates, Jr.,
  ``Why auxiliary fields matter: The strange case of the 4-D, N=1 supersymmetric QCD effective action. 2.,''
  Nucl.\ Phys.\ B {\bf 485}, 145 (1997)
  [hep-th/9606109].
  %%CITATION = HEP-TH/9606109;%%
  %32 citations counted in INSPIRE as of 21 Jun 2014

\bibitem{Nitta:2001rh} 
  M.~Nitta,
  ``A Note on supersymmetric WZW term in four dimensions,''
  Mod.\ Phys.\ Lett.\ A {\bf 15}, 2327 (2000)
  [hep-th/0101166].
  %%CITATION = HEP-TH/0101166;%%
  %3 citations counted in INSPIRE as of 21 Jun 2014

\bibitem{Nemeschansky:1984cd} 
  D.~Nemeschansky and R.~Rohm,
  ``Anomaly Constraints On Supersymmetric Effective Lagrangians,''
  Nucl.\ Phys.\ B {\bf 249}, 157 (1985).
  %%CITATION = NUPHA,B249,157;%%
  %31 citations counted in INSPIRE as of 21 Jun 2014

\bibitem{Gates:2000rp} 
  S.~J.~Gates, Jr., M.~T.~Grisaru, M.~E.~Knutt and S.~Penati,
  ``The Superspace WZNW action for 4-D, N=1 supersymmetric QCD,''
  Phys.\ Lett.\ B {\bf 503}, 349 (2001)
  [hep-ph/0012301];
  %%CITATION = HEP-PH/0012301;%%
  %3 citations counted in INSPIRE as of 22 Jun 2014
  S.~J.~Gates, Jr., M.~T.~Grisaru, M.~E.~Knutt, S.~Penati and H.~Suzuki,
  ``Supersymmetric gauge anomaly with general homotopic paths,''
  Nucl.\ Phys.\ B {\bf 596}, 315 (2001)
  [hep-th/0009192];
  %%CITATION = HEP-TH/0009192;%%
  %11 citations counted in INSPIRE as of 22 Jun 2014
  S.~J.~Gates, Jr., M.~T.~Grisaru and S.~Penati,
  ``Holomorphy, minimal homotopy and the 4-D, N=1 supersymmetric Bardeen-Gross-Jackiw anomaly,''
  Phys.\ Lett.\ B {\bf 481}, 397 (2000)
  [hep-th/0002045].
  %%CITATION = HEP-TH/0002045;%%
  %9 citations counted in INSPIRE as of 22 Jun 2014

\bibitem{RoTs}
  M.~Rocek and A.~A.~Tseytlin,
  ``Partial breaking of global D = 4 supersymmetry, constrained superfields, and three-brane actions,''
  Phys.\ Rev.\ D {\bf 59} (1999) 106001
  [hep-th/9811232].
  %%CITATION = HEP-TH/9811232;%%
  %114 citations counted in INSPIRE as of 03 Jun 2014

\bibitem{BeNeSc}
  E.~A.~Bergshoeff, R.~I.~Nepomechie and H.~J.~Schnitzer,
  ``Supersymmetric Skyrmions in Four-dimensions,''
  Nucl.\ Phys.\ B {\bf 249} (1985) 93.
  %%CITATION = NUPHA,B249,93;%%
  %33 citations counted in INSPIRE as of 14 Feb 2014

\bibitem{Fr}
  L.~Freyhult,
  ``The Supersymmetric extension of the Faddeev model,''
  Nucl.\ Phys.\ B {\bf 681} (2004) 65
  [hep-th/0310261].
  %%CITATION = HEP-TH/0310261;%%
  %11 citations counted in INSPIRE as of 17 Jun 2014

\bibitem{Adam:2011hj} 
  C.~Adam, J.~M.~Queiruga, J.~Sanchez-Guillen and A.~Wereszczynski,
  ``N=1 supersymmetric extension of the baby Skyrme model,''
  Phys.\ Rev.\ D {\bf 84}, 025008 (2011)
  [arXiv:1105.1168 [hep-th]].
  %%CITATION = ARXIV:1105.1168;%%
  %13 citations counted in INSPIRE as of 21 Jun 2014

\bibitem{AdQuSaGuWe}
  C.~Adam, J.~M.~Queiruga, J.~Sanchez-Guillen and A.~Wereszczynski,
  ``Extended Supersymmetry and BPS solutions in baby Skyrme models,''
  JHEP {\bf 1305} (2013) 108
  [arXiv:1304.0774 [hep-th]].
  %%CITATION = ARXIV:1304.0774;%%
  %2 citations counted in INSPIRE as of 24 May 2014

\bibitem{AdQuSaGuWe2}
  C.~Adam, J.~M.~Queiruga, J.~Sanchez-Guillen and A.~Wereszczynski,
  ``BPS bounds in supersymmetric extensions of K field theories,''
  Phys.\ Rev.\ D {\bf 86} (2012) 105009
  [arXiv:1209.6060 [hep-th]].
  %%CITATION = ARXIV:1209.6060;%%
  %5 citations counted in INSPIRE as of 03 Jun 2014

\bibitem{AdQuSaGuWe3}
  C.~Adam, J.~M.~Queiruga, J.~Sanchez-Guillen and A.~Wereszczynski,
  ``Supersymmetric K field theories and defect structures,''
  Phys.\ Rev.\ D {\bf 84} (2011) 065032
  [arXiv:1107.4370 [hep-th]].
  %%CITATION = ARXIV:1107.4370;%%
  %13 citations counted in INSPIRE as of 15 Jun 2014

\bibitem{KhLeOv}
  J.~Khoury, J.~-L.~Lehners and B.~Ovrut,
  ``Supersymmetric P(X,$\phi$) and the Ghost Condensate,''
  Phys.\ Rev.\ D {\bf 83} (2011) 125031
  [arXiv:1012.3748 [hep-th]].
  %%CITATION = ARXIV:1012.3748;%%
  %33 citations counted in INSPIRE as of 24 May 2014

%\cite{Buchbinder:1994iw}
\bibitem{Buchbinder:1994iw} 
  I.~L.~Buchbinder, S.~Kuzenko and Z.~Yarevskaya,
  ``Supersymmetric effective potential: Superfield approach,''
  Nucl.\ Phys.\ B {\bf 411}, 665 (1994);
  %%CITATION = NUPHA,B411,665;%%
  %66 citations counted in INSPIRE as of 22 Aug 2014
%\cite{Buchbinder:1998qv}
%\bibitem{Buchbinder:1998qv} 
  I.~L.~Buchbinder and S.~M.~Kuzenko,
  ``Ideas and methods of supersymmetry and supergravity: Or a walk through superspace,''
  Bristol, UK: IOP (1998) 656 p; 
  %14 citations counted in INSPIRE as of 22 Aug 2014
%\cite{Kuzenko:2014ypa}
%\bibitem{Kuzenko:2014ypa} 
  S.~M.~Kuzenko and S.~J.~Tyler,
  ``The one-loop effective potential of the Wess-Zumino model revisited,''
  arXiv:1407.5270 [hep-th].
  %%CITATION = ARXIV:1407.5270;%%

\bibitem{Banin:2006db} 
  A.~T.~Banin, I.~L.~Buchbinder and N.~G.~Pletnev,
  ``On quantum properties of the four-dimensional generic chiral superfield model,''
  Phys.\ Rev.\ D {\bf 74}, 045010 (2006)
  [hep-th/0606242].
  %%CITATION = HEP-TH/0606242;%%
  %6 citations counted in INSPIRE as of 21 Jun 2014

\bibitem{SaYaYo}
  S.~Sasaki, M.~Yamaguchi and D.~Yokoyama,
  ``Supersymmetric DBI inflation,''
  Phys.\ Lett.\ B {\bf 718} (2012) 1
  [arXiv:1205.1353 [hep-th]].
  %%CITATION = ARXIV:1205.1353;%%
  %7 citations counted in INSPIRE as of 24 May 2014

\bibitem{KoLeOv}
  M.~Koehn, J.~-L.~Lehners and B.~A.~Ovrut,
  ``Higher-Derivative Chiral Superfield Actions Coupled to N=1 Supergravity,''
  Phys.\ Rev.\ D {\bf 86} (2012) 085019
  [arXiv:1207.3798 [hep-th]].
  %%CITATION = ARXIV:1207.3798;%%
  %19 citations counted in INSPIRE as of 24 May 2014

\bibitem{FaKe}
  F.~Farakos and A.~Kehagias,
  ``Emerging Potentials in Higher-Derivative Gauged Chiral Models Coupled to N=1 Supergravity,''
  JHEP {\bf 1211} (2012) 077
  [arXiv:1207.4767 [hep-th]].
  %%CITATION = ARXIV:1207.4767;%%
  %14 citations counted in INSPIRE as of 24 May 2014

\bibitem{KhLeOv2}
  J.~Khoury, J.~-L.~Lehners and B.~A.~Ovrut,
  ``Supersymmetric Galileons,''
  Phys.\ Rev.\ D {\bf 84} (2011) 043521
  [arXiv:1103.0003 [hep-th]].
  %%CITATION = ARXIV:1103.0003;%%
  %56 citations counted in INSPIRE as of 03 Jun 2014

\bibitem{KoLeOv2}
  M.~Koehn, J.~-L.~Lehners and B.~Ovrut,
  ``Ghost condensate in $N=1$ supergravity,''
  Phys.\ Rev.\ D {\bf 87} (2013),  065022
  [arXiv:1212.2185 [hep-th]].
  %%CITATION = ARXIV:1212.2185;%%
  %9 citations counted in INSPIRE as of 03 Jun 2014

%\cite{Nitta:2014pwa}
\bibitem{Nitta:2014pwa} 
  M.~Nitta and S.~Sasaki,
  ``BPS States in Supersymmetric Chiral Models with Higher Derivative
		Terms,''
  Phys.\ Rev.\ D {\bf 90} (2014), 105001 
  arXiv:1406.7647 [hep-th].
  %%CITATION = ARXIV:1406.7647;%%
  %1 citations counted in INSPIRE as of 02 Aug 2014

%\cite{Bolognesi:2014ova}
\bibitem{Bolognesi:2014ova} 
  S.~Bolognesi and W.~Zakrzewski,
  ``Baby Skyrme Model, Near-BPS Approximations and Supersymmetric Extensions,''
  arXiv:1407.3140 [hep-th].
  %%CITATION = ARXIV:1407.3140;%%

\bibitem{Wess:1992cp} 
  J.~Wess and J.~Bagger,
  ``Supersymmetry and supergravity,''
  Princeton, USA: Univ. Pr. (1992) 259 p
  %48 citations counted in INSPIRE as of 22 Jun 2014


%%%%%%%%%%%%%%%%%%%%%%%%%%%%%%%%%

%\cite{BarMoshe:1994rx}
\bibitem{BarMoshe:1994rx} 
  D.~Bar-Moshe and M.~S.~Marinov,
  ``Realization of compact Lie algebras in Kahler manifolds,''
  J.\ Phys.\ A {\bf 27}, 6287 (1994)
  [hep-th/9407092].
  %%CITATION = HEP-TH/9407092;%%
  %10 citations counted in INSPIRE as of 02 Aug 2014

\bibitem{Borel:1954}
A.~Borel,
Proc. Nat. Acad. Sci. {\bf 40} (1954) 1147.

%\cite{Aoyama:2000vb}
\bibitem{Aoyama:2000vb} 
  S.~Aoyama,
  ``The Four Fermi coupling of the supersymmetric nonlinear sigma model on G / S$\times${U(1)}$^k$,''
  Nucl.\ Phys.\ B {\bf 578}, 449 (2000)
  [hep-th/0001160].
  %%CITATION = HEP-TH/0001160;%%
  %12 citations counted in INSPIRE as of 29 Jul 2014

%%%%%%%%%%%%%%%%%%%%%%%%%%%%%%%%%%%%%%%%%%%

%\cite{Higashijima:2000wz}
\bibitem{Higashijima:2000wz} 
  K.~Higashijima and M.~Nitta,
  ``Kahler normal coordinate expansion in supersymmetric theories,''
  Prog.\ Theor.\ Phys.\  {\bf 105}, 243 (2001)
  [hep-th/0006027];
  %%CITATION = HEP-TH/0006027;%%
  %33 citations counted in INSPIRE as of 12 Aug 2014
%\cite{Higashijima:2002fq}
%\bibitem{Higashijima:2002fq} 
  K.~Higashijima, E.~Itou and M.~Nitta,
  ``Normal coordinates in Kahler manifolds and the background field method,''
  Prog.\ Theor.\ Phys.\  {\bf 108}, 185 (2002)
  [hep-th/0203081].
  %%CITATION = HEP-TH/0203081;%%
  %24 citations counted in INSPIRE as of 12 Aug 2014

\bibitem{Adam:2010fg} 
  C.~Adam, J.~Sanchez-Guillen and A.~Wereszczynski,
  ``A Skyrme-type proposal for baryonic matter,''
  Phys.\ Lett.\ B {\bf 691}, 105 (2010)
  [arXiv:1001.4544 [hep-th]];
  %%CITATION = ARXIV:1001.4544;%%
  %\bibitem{Adam:2010ds} 
  C.~Adam, J.~Sanchez-Guillen and A.~Wereszczynski,
  ``A BPS Skyrme model and baryons at large $N_c$,''
  Phys.\ Rev.\ D {\bf 82}, 085015 (2010)
  [arXiv:1007.1567 [hep-th]].
  %%CITATION = ARXIV:1007.1567;%%



\bibitem{Hanany:2003hp} 
  A.~Hanany and D.~Tong,
  ``Vortices, instantons and branes,''
  JHEP {\bf 0307}, 037 (2003)
  [hep-th/0306150];
  %%CITATION = HEP-TH/0306150;%%
  %290 citations counted in INSPIRE as of 22 Jun 2014
  R.~Auzzi, S.~Bolognesi, J.~Evslin, K.~Konishi and A.~Yung,
  ``NonAbelian superconductors: Vortices and confinement in N=2 SQCD,''
  Nucl.\ Phys.\ B {\bf 673}, 187 (2003)
  [hep-th/0307287];
  %%CITATION = HEP-TH/0307287;%%
  %255 citations counted in INSPIRE as of 22 Jun 2014
  M.~Eto, Y.~Isozumi, M.~Nitta, K.~Ohashi and N.~Sakai,
  ``Moduli space of non-Abelian vortices,''
  Phys.\ Rev.\ Lett.\  {\bf 96}, 161601 (2006)
  [hep-th/0511088];
  %%CITATION = HEP-TH/0511088;%%
  %139 citations counted in INSPIRE as of 22 Jun 2014
  M.~Eto, K.~Konishi, G.~Marmorini, M.~Nitta, K.~Ohashi, W.~Vinci and N.~Yokoi,
  ``Non-Abelian Vortices of Higher Winding Numbers,''
  Phys.\ Rev.\ D {\bf 74}, 065021 (2006)
  [hep-th/0607070].
  %%CITATION = HEP-TH/0607070;%%
  %89 citations counted in INSPIRE as of 22 Jun 2014

\bibitem{Eto:2012qda} 
  M.~Eto, T.~Fujimori, M.~Nitta, K.~Ohashi and N.~Sakai,
  ``Higher Derivative Corrections to Non-Abelian Vortex Effective Theory,''
  Prog.\ Theor.\ Phys.\  {\bf 128}, 67 (2012)
  [arXiv:1204.0773 [hep-th]].
  %%CITATION = ARXIV:1204.0773;%%
  %9 citations counted in INSPIRE as of 21 Jun 2014

\bibitem{Eto:2004rz} 
  M.~Eto, Y.~Isozumi, M.~Nitta, K.~Ohashi and N.~Sakai,
  ``Instantons in the Higgs phase,''
  Phys.\ Rev.\ D {\bf 72}, 025011 (2005)
  [hep-th/0412048];
  %%CITATION = HEP-TH/0412048;%%
  %82 citations counted in INSPIRE as of 22 Jun 2014
  T.~Fujimori, M.~Nitta, K.~Ohta, N.~Sakai and M.~Yamazaki,
  ``Intersecting Solitons, Amoeba and Tropical Geometry,''
  Phys.\ Rev.\ D {\bf 78}, 105004 (2008)
  [arXiv:0805.1194 [hep-th]].
  %%CITATION = ARXIV:0805.1194;%%
  %22 citations counted in INSPIRE as of 22 Jun 2014

%\cite{Tong:2005un}
\bibitem{Tong:2005un} 
  D.~Tong,
  ``TASI lectures on solitons: Instantons, monopoles, vortices and kinks,''
  hep-th/0509216.
  %%CITATION = HEP-TH/0509216;%%
  %177 citations counted in INSPIRE as of 18 Aug 2014


%\cite{Eto:2006pg}
\bibitem{Eto:2006pg} 
  M.~Eto, Y.~Isozumi, M.~Nitta, K.~Ohashi and N.~Sakai,
  ``Solitons in the Higgs phase: The Moduli matrix approach,''
  J.\ Phys.\ A {\bf 39}, R315 (2006)
  [hep-th/0602170].
  %%CITATION = HEP-TH/0602170;%%
  %202 citations counted in INSPIRE as of 17 Aug 2014
%\cite{Eto:2005sw}
%\bibitem{Eto:2005sw} 
  M.~Eto, Y.~Isozumi, M.~Nitta and K.~Ohashi,
  ``1/2, 1/4 and 1/8 BPS equations in SUSY Yang-Mills-Higgs systems: Field theoretical brane configurations,''
  Nucl.\ Phys.\ B {\bf 752}, 140 (2006)
  [hep-th/0506257].
  %%CITATION = HEP-TH/0506257;%%
  %53 citations counted in INSPIRE as of 18 Aug 2014

%\cite{Shifman:2007ce}
\bibitem{Shifman:2007ce} 
  M.~Shifman and A.~Yung,
  ``Supersymmetric Solitons and How They Help Us Understand Non-Abelian Gauge Theories,''
  Rev.\ Mod.\ Phys.\  {\bf 79}, 1139 (2007)
  [hep-th/0703267];
  %%CITATION = HEP-TH/0703267;%%
  %151 citations counted in INSPIRE as of 17 Aug 2014
%\cite{Shifman:2009zz}
%\bibitem{Shifman:2009zz} 
  M.~Shifman and A.~Yung,
  ``Supersymmetric solitons,''
  Cambridge, UK: Cambridge Univ. Pr. (2009) 259 p
  %6 citations counted in INSPIRE as of 18 Aug 2014

%\cite{Eto:2008yi}
\bibitem{Eto:2008yi} 
  M.~Eto, T.~Fujimori, S.~B.~Gudnason, K.~Konishi, M.~Nitta, K.~Ohashi and W.~Vinci,
  ``Constructing Non-Abelian Vortices with Arbitrary Gauge Groups,''
  Phys.\ Lett.\ B {\bf 669}, 98 (2008)
  [arXiv:0802.1020 [hep-th]].
  %%CITATION = ARXIV:0802.1020;%%
  %52 citations counted in INSPIRE as of 03 Aug 2014

%\cite{Eto:2009bg}
\bibitem{Eto:2009bg} 
  M.~Eto, T.~Fujimori, S.~B.~Gudnason, K.~Konishi, T.~Nagashima, M.~Nitta, K.~Ohashi and W.~Vinci,
  ``Non-Abelian Vortices in SO(N) and USp(N) Gauge Theories,''
  JHEP {\bf 0906}, 004 (2009)
  [arXiv:0903.4471 [hep-th]];
  %%CITATION = ARXIV:0903.4471;%%
  %29 citations counted in INSPIRE as of 03 Aug 2014
%\cite{Ferretti:2007rp}
%\bibitem{Ferretti:2007rp} 
  L.~Ferretti, S.~B.~Gudnason and K.~Konishi,
  ``Non-Abelian vortices and monopoles in SO(N) theories,''
  Nucl.\ Phys.\ B {\bf 789}, 84 (2008)
  [arXiv:0706.3854 [hep-th]];
  %%CITATION = ARXIV:0706.3854;%%
  %32 citations counted in INSPIRE as of 03 Aug 2014
%\cite{Eto:2008qw}
%\bibitem{Eto:2008qw} 
  M.~Eto, T.~Fujimori, S.~B.~Gudnason, M.~Nitta and K.~Ohashi,
  ``SO and US(p) Kahler and Hyper-Kahler Quotients and Lumps,''
  Nucl.\ Phys.\ B {\bf 815}, 495 (2009)
  [arXiv:0809.2014 [hep-th]]; 
  %%CITATION = ARXIV:0809.2014;%%
  %26 citations counted in INSPIRE as of 03 Aug 2014
%\cite{Eto:2011cv}
%\bibitem{Eto:2011cv} 
  M.~Eto, T.~Fujimori, S.~B.~Gudnason, Y.~Jiang, K.~Konishi, M.~Nitta and K.~Ohashi,
  ``Vortices and Monopoles in Mass-deformed SO and USp Gauge Theories,''
  JHEP {\bf 1112}, 017 (2011)
  [arXiv:1108.6124 [hep-th]].
  %%CITATION = ARXIV:1108.6124;%%
  %10 citations counted in INSPIRE as of 03 Aug 2014


\bibitem{Shifman:2003uh} 
  M.~Shifman and A.~Yung,
  ``Localization of nonAbelian gauge fields on domain walls at weak coupling (D-brane prototypes II),''
  Phys.\ Rev.\ D {\bf 70}, 025013 (2004)
  [hep-th/0312257];
  %%CITATION = HEP-TH/0312257;%%
  %77 citations counted in INSPIRE as of 22 Jun 2014
  M.~Eto, T.~Fujimori, M.~Nitta, K.~Ohashi and N.~Sakai,
  ``Domain walls with non-Abelian clouds,''
  Phys.\ Rev.\ D {\bf 77}, 125008 (2008)
  [arXiv:0802.3135 [hep-th]].
  %%CITATION = ARXIV:0802.3135;%%
  %22 citations counted in INSPIRE as of 22 Jun 2014

\bibitem{Eto:2005cc} 
  M.~Eto, M.~Nitta, K.~Ohashi and D.~Tong,
  ``Skyrmions from instantons inside domain walls,''
  Phys.\ Rev.\ Lett.\  {\bf 95}, 252003 (2005)
  [hep-th/0508130].
  %%CITATION = HEP-TH/0508130;%%
  %31 citations counted in INSPIRE as of 22 Jun 2014


\bibitem{Eto:2006uw} 
  M.~Eto, Y.~Isozumi, M.~Nitta, K.~Ohashi and N.~Sakai,
  ``Manifestly supersymmetric effective Lagrangians on BPS solitons,''
  Phys.\ Rev.\ D {\bf 73}, 125008 (2006)
  [hep-th/0602289].
  %%CITATION = HEP-TH/0602289;%%
  %64 citations counted in INSPIRE as of 21 Jun 2014

\end{thebibliography}
\end{document}